\documentclass[%
 reprint,
 amsmath,amssymb,
 aps,
]{revtex4-1}

\usepackage{graphicx}%
\usepackage{dcolumn}%
\usepackage{bm}%

\usepackage[colorlinks = true, %Turn the links in table of contents into clickable links
            linkcolor = blue,
            urlcolor  = blue,
            citecolor = blue,
            anchorcolor = blue]{hyperref}

\usepackage{braket}
\usepackage{xcolor}
\usepackage{soul} 
\usepackage{ulem}

\begin{document}

\preprint{APS/123-QED}

\title{Topological phonon analysis of the 2D buckled honeycomb lattice: an application to real materials}%

\author{Martin Gutierrez-Amigo$^{1,2}$}
 \email{mgutierrez042@ikasle.ehu.eus}
\author{Maia G. Vergniory$^{3,4}$}%
\author{Ion Errea$^{2,3,5}$}
\author{J. L. Ma\~nes$^1$}
 \email{wmpmapaj@lg.ehu.es}

\affiliation{$^1$Departamento de F\'isica, Facultad de Ciencia y Tecnolog\'ia, Universidad del País Vasco (UPV/EHU), Apartado 644, 48080 Bilbao, Spain} 
\affiliation{$^2$Centro de F\'isica de Materiales (CSIC-UPV/EHU), Manuel de Lardizabal pasealekua 5, 20018 Donostia/San Sebasti\'an, Spain}
\affiliation{$^3$Donostia International Physics Center (DIPC), Manuel Lardizabal pasealekua 4, 20018 Donostia/San Sebasti\'an, Spain}
\affiliation{$^4$ Max Planck for Chemical Physics of Solids, Noethnitzer Str. 40, 01187 Dresden, Germany}
\affiliation{$^5$Fisika Aplikatua Saila, Gipuzkoako Ingeniaritza Eskola, University of the Basque Country (UPV/EHU), Europa Plaza 1, 20018 Donostia/San Sebasti\'an, Spain}

\date{\today}%

\begin{abstract}
By means of group theory, topological quantum chemistry, first-principles and Monte Carlo calculations, we analyze the topology of the 2D buckled honeycomb lattice phonon spectra. Taking the pure crystal structure as an input, we show that eleven distinct phases are possible, five of which necessarily have non-trivial topology according to topological quantum chemistry. Another four of them are also identified as topological using Wilson loops in an analytical model that includes all the symmetry allowed force constants up to third nearest neighbors, making a total of nine topological phases. We then compute the \textit{ab initio} phonon spectra  for the two-dimensional crystals of  Si, Ge, P,  As and Sb in this structure and construct its phase diagram. Despite the large proportion of topological phases found in the analytical model, all of the crystals lie in a trivial phase. By analyzing the force constants space using Monte Carlo calculations, we elucidate why topological phonon phases are physically difficult to realize in real materials with this crystal structure.
\end{abstract}

\maketitle

\section{\label{sec_intro}Introduction}

Topological materials are known for having exotic electronic properties such as symmetry protected surface states, edge states or unconventional electromagnetic activity \cite{topological_surface_states,edge_states}. Following the first topological insulator realization \cite{first_ins} in 2007, the concept of symmetry protected topological (SPT) orders  \cite{SPT1,SPT2,SPT3} was extended to all sorts of symmetry settings with the advent of topological crystalline insulators \cite{TCI1,TCI2,TCI3,TCI4,TCI5,TCI6,TCI7,TCI8,TCI10,TCI11,TCI12}. However, with the exception of the Fu-Kane parity criterion \cite{Fu-Kane_parity_criterion} and a few other symmetry based diagnosis methods \cite{sym_methods1,sym_methods2,sym_methods3}, there was not a clear link between symmetry and topology for general symmetry settings. As a result, the calculation of most topological band invariants used in the  prediction and diagnosis of topological materials had to be carried out  numerically with computationally expensive \textit{ab initio} methods, and the rate of discovery of new materials was consequently rather slow.

Recently, much more  powerful links have been established between the topology of the electronic spectrum  and the crystal symmetry thanks to the theory of symmetry indicators of band topology \cite{sym_indicators1,sym_indicators2},  band combinatorics \cite{band_combinatorics}, and topological quantum chemistry (TQC) \cite{TQC}. These formalisms have provided a reliable and systematic way to search for all  the topologically non-trivial phases compatible with a given crystal structure. This has led to the discovery of  thousands of materials with non-trivial electronic topology, showing that the existence  of topological electron bands,  previously considered a rarity, is rather frequent in nature.
In particular, the systematic application of the methods of TQC have enormously enlarged the number of known topological materials \cite{Vergniory:2019qy, all_bands} and led to new and more refined methods of classifying  their topology \cite{disconnected_EBR,Bouhon_2019, WilsonLoops, wannier_obstructions}.

As phonons are behind many important properties of solids, such as transport, optical and thermal responses, and superconductivity, finding materials with topological phonon bands is likely to  have a revolutionary impact on solid state physics. The search for non-trivial phonon topology, which has been based on more traditional  methods \cite{top1,top2,top3,top4,top5,top6,top7,top8,top9,top10}, has proceeded at a very slow pace, focusing on degeneracies with topological charge, such  as Weyl points, high degenerate Weyls, and nodal lines and rings \cite{top_phonons_perovskites,top_phonons_data_driven}. Recently, a gigantic step has been taken and a phonon catalogue has been launched \cite{phonon_catalogue} applying TQC to 3D materials. In that work they point out the almost absence of fragile cumulative topology for phonons. Inspired by these results, we deepen in the phonon study by performing Monte Carlo calculations and analysing the force constants in the 2D buckled honeycomb lattice.

The TQC analysis relies on detecting an obstruction to a localized real space interpretation of isolated subsets of phonon bands. Whenever this obstruction is present, the subset has non-trivial topology. In some cases this obstruction can be diagnosed just from the irreducible representations (irreps) describing how those bands transform at the high symmetry points (HSPs) in the Brillouin zone. In practice, this analysis consists of three steps. First, one finds the irreps describing how phonons transform at the HSPs. Second, a compatibility problem is solved, in which one tries to connect the bands forming isolated subsets separated by gaps in ways that are consistent with the system symmetries and the existence of the acoustic zero frequency modes. Each of these band configurations constitutes a phase. Finally, if the irreps of an isolated subset cannot be obtained from the sum of elementary band representations (EBRs), the corresponding phase is necessarily topological. Note that even if the irreps of all the isolated subsets in a given phase can be obtained from the sum of elementary band representations, the phase could still be topological. Thus, an extra step in order to fully diagnose the topology involves the construction of an analytical model that reproduces the different phases and the computation of Wilson loop spectra for the different phases. 

In this paper we extend the work in \cite{graphene_phonons}, where TQC methods were used to find and characterize four new topological phases for phonons on the planar honeycomb lattice, to the analysis of the buckled honeycomb lattice (BHL). This  is an important  system since the planar honeycomb lattice is unstable for atoms larger than carbon and two-dimensional materials based on Si, Ge, P,  As, and Sb crystallize in  the buckled honeycomb lattice.
As we will see, this introduces additional complications due to couplings between in- and off- plane modes, which decouple in the planar limit,  and gives rise to a whole array of phases not present on the planar honeycomb. Finally, we compute the density functional perturbation theory (DFPT) \cite{RevModPhys.73.515}  phonon spectra  for several monoatomic crystals with the buckled honeycomb structure and place  them in the phase diagram.  We also  present a Monte Carlo analysis of the space of force constants that explains why topological phonon phases are physically difficult to realize in real materials even in the presence of applied strain.

This paper is organized as follows. 
A discussion on how the TQC machinery is adapted to the study of phonon spectra of the BHL is presented in Section \ref{sec_TQC}. 
An analytical model that includes all the symmetry compatible couplings up to third nearest neighbors is constructed  in Section \ref{sec_analytic_model}. The model is used to compute  Wilson loop spectra and fully classify the topology of all the phases obtained in Section \ref{sec_TQC}. In Section \ref{sec_DFPT} we use Quantum Espresso \cite{giannozzi2009quantum,QE-2017} to compute the phonon spectra of  real materials and locate them on the phase diagram. Moreover, we show how the analytical model can be used to study the phase diagram by means of  a Monte Carlo method. Finally, the summary and conclusions are presented in Section \ref{sec_conclusions}.

\section{\label{sec_TQC}
Topological quantum chemistry application to phonons of  the buckled honeycomb lattice}

\subsection{\label{BRs_elec_phon} Band representations for electrons and phonons}

In electronic systems  a band representation (BR) \hbox{\cite{zak_sym,zak_bandrep,EBRs}} can be understood as a mathematical construction that links the real space orbital description to the reciprocal space momentum picture. 
More concretely, given a crystal with a set of orbitals  closed under the action of the space group  $G$ of the crystal, the transformation properties of  the orbitals under $G$ define a band representation. One says that the band representation is induced by the set of orbitals.  Note that, in order to be closed under the translations in $G$, the set must contain infinitely many orbitals and band representations are always infinite-dimensional, which is at  the origin of some  counter-intuitive properties. 

Although orbitals are localized objects and band representations are initially defined in real space, we can always take Bloch-like combinations of orbitals with well defined crystal momentum, which amounts to a simple change of basis, and get a description  in reciprocal space. In practice, this means that each band representation induces a collection of little group irreps at every point  in the Brillouin zone  \cite{TQC}. Although this collection of irreps can be considered as a footprint of the band representation, it is important to bear in mind  that the band representation is not uniquely specified by its footprint, as different (inequivalent) band representations, and even representations that are not BRs,  can give rise to identical sets of irreps at all the points in the Brillouin zone \cite{PhysRevB.105.125115}. The reason is that the set of irreps at all the points in the Brillouin zone does  not exhaust all the information contained in the band representation. This phenomenon has no analogue in the  case of  ordinary, finite-dimensional representations and has  important consequences for the irrep-based detection of topological phases.

Elementary band representations (EBRs), as introduced in the TQC  \cite{TQC} formalism, are  induced from a set of orbitals that transform under an irreducible representation $D$ of the local site symmetry group $G_q$ of a maximal symmetry Wyckoff position $q$ (with some exceptions \cite{MICHEL2001377,disconnected_EBR,EBRs}). Band representations and EBRs are related by the fact that a band representation is either an EBR or can be written as a sum of EBRs. Whenever a subset of bands that is separated by a gap from the rest of the bands does not transform as a band representation, the subset does not have an atomic limit and is topologically non-trivial. Therefore, if  a subset of bands cannot be written as a sum of EBRs, it cannot transform as a band representation and must be topological. In practical terms, if the irreps of an isolated subset of bands can not  be induced from any sum of EBRs, we can conclude that the subset does not transform as a band representation and is topologically non-trivial. This test is easily implemented and yields a practical method to identify topological phases in electronic systems. Note, however,  that even when  the irreps of an isolated subset of bands \textit{can} be obtained from the sum of EBRs, this does not guaratee that the subset transforms as a band representation and the phase could still be topological~\cite{EBRs}.  We will return to this important point in the next Section.

The concept of inducing a band representation can be easily extended to the phononic case \cite{Zak_phonons,graphene_phonons}. Instead of having orbitals or Wannier  functions (WFs) as a basis, one has a set of vectors describing the displacements located on every atom in the crystal. These displacements (real space) transform under the vector representation $V(g)$ for symmetry elements $g$  of the site symmetry group $G_q$. As phonons are a combination of local displacements, one can induce a band representation describing how phonons (reciprocal space) transform under the full space group symmetry operations from the vector representation of one of the site symmetry groups of the occupied Wyckoff positions (WPs).

When inducing the band representation from the displacements centered on the atoms, we obtain \textit{all} the phononic bands for the crystal. But this should not be mistaken with the possibility of inducing connected subsets of bands from `Wannier like" functions for phonons, even when these WFs are not centered on atomic positions. The matter of defining a localized basis for vibrations was already discussed by W. Kohn \cite{Wannier_phonons} and more recently in terms of the position operator in Ref. \cite{Wannier_phonons_pos}. This possibility generalizes the concept of EBRs to phonons in an equivalent way to the electronic case, where EBRs are induced from WFs that transform under an irreducible representation of the local site symmetry group $G_q$ of any maximal symmetry Wyckoff position.

\begin{figure}[t]
\includegraphics[width=9cm,keepaspectratio]{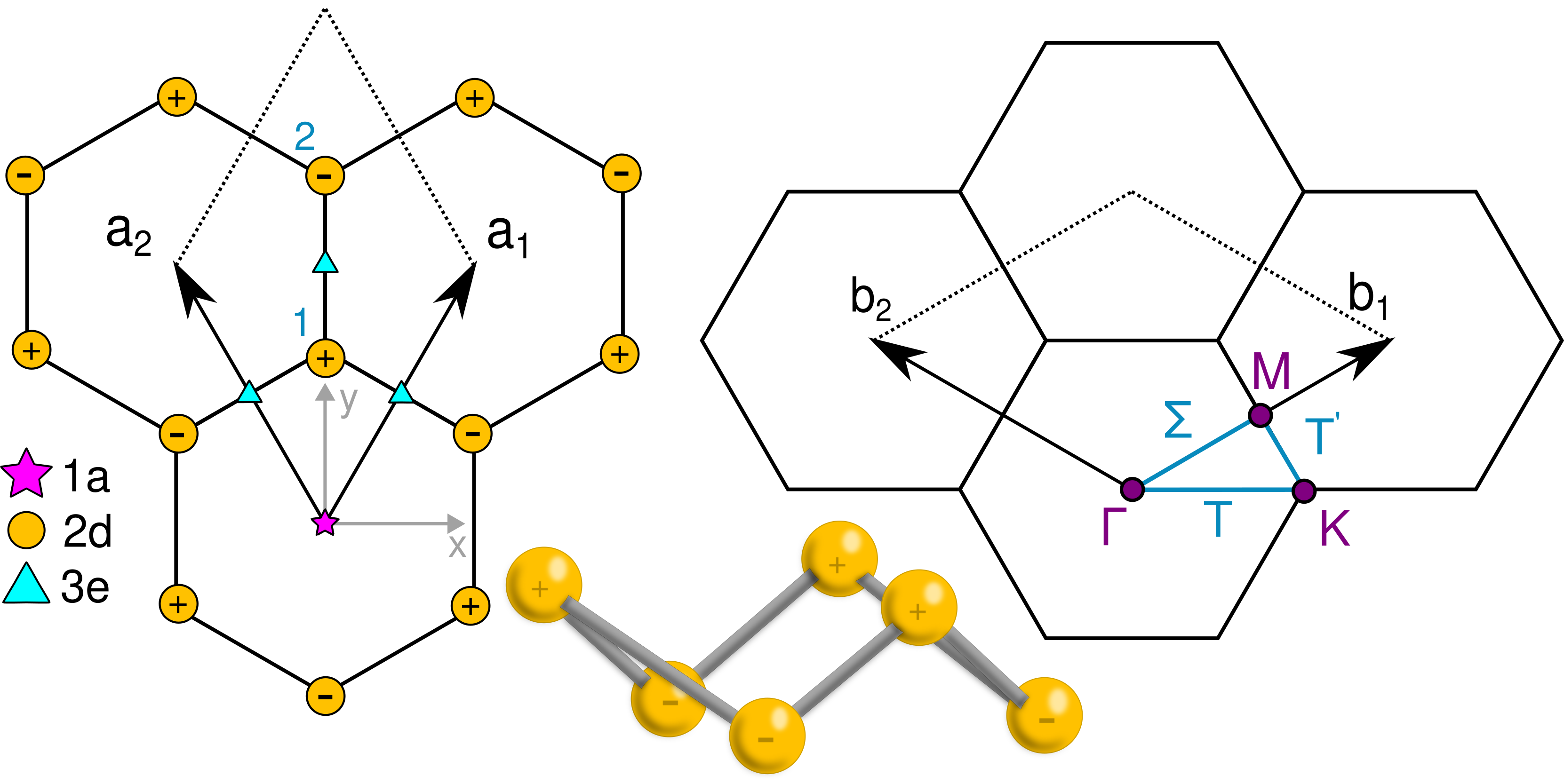}
\caption{\label{fig_vectors} Buckled honeycomb lattice with real an reciprocal space lattice vectors. The structure is formed with two atoms per cell, placed at the $2d$ maximal Wyckoff positions, each displaced along the $z$ axis in opposite directions. The little groups for the $\Gamma$, K, M points and T, T', $\Sigma$ lines are $D_{3d}$, $D_3$, $C_{2h}$ and $C_2$, $C_2$, $C_{1h}$, respectively.}
\end{figure}

\subsection{\label{Mechanical_BR} Mechanical BR for the BHL structure}
As shown in Fig.~\ref{fig_vectors} the BHL is a 2D structure that can be obtained by giving opposite vertical (perpendicular to the sample plane)  displacements to the two atoms in the primitive cell of the planar honeycomb lattice. These displacements 
break the mirror symmetry with respect to the  plane of the sample  and are thus responsible for the couplings between off- and in-plane phonons, which are forbidden for the planar honeycomb. The layer group leaving the BHL invariant is  LG 72 ($p\overline{3}m1$), which corresponds to the space group SG 164 ($P\overline{3}m1$) with point group $D_{3d}$ ($\overline{3}m$). 
The direct and reciprocal lattice vectors satisfying $\mathbf{a}_i \cdot \mathbf{b}_j = 2 \pi \delta_{ij} $ are shown in Fig \ref{fig_vectors}. In this basis, the high symmetry points are located at $\Gamma=(0,0)$, $K=\frac{1}{3}(\textbf{b}_1-\textbf{b}_2)$ and $M=\frac{\textbf{b}_1}{2}$, while the atoms are located at the Wyckoff position $2d$ with site-symmetry group $C_{3v}$.

As mentioned above, the phononic BR for the crystal, also known as the mechanical BR,  is induced from the vector representations of the atomic site symmetry groups
\begin{equation}
\rho_G=\bigoplus_q V_{G_q} \uparrow G,
\end{equation}
where  $q$ runs over all occupied Wyckoff positions. By decomposing  the vector representation  into irreducible representations $D^{(i)}_{G_q}$,  the mechanical BR can be written as a sum of EBRs
\begin{equation}\label{MBR}
\rho_G=\bigoplus_q V_{G_q} \uparrow G=\bigoplus_{q,i} D^{(i)}_{G_q} \uparrow G
\end{equation}
In our case, the vector representation of $C_{3v}$ is reducible, \hbox{$V=A_1+E$}, and according to Eq.~(\ref{MBR}) the mechanical BR is given by 
\begin{equation}\label{MBRebrs}
M=V|_{2d} \uparrow G = A_1|_{2d} \uparrow G \oplus E|_{2d} \uparrow G. 
\end{equation}
In other words, the mechanical BR can be written as the sum of the EBRs induced from $A_1$ and $E$ at the WP $2d$.
We can avoid the actual computation of the mechanical BR by noting that the little group irreps induced by any EBR at any point $\mathbf{k}$  in the BZ are given by the \hbox{\textit{BandRep}} application at the Bilbao Christallographic Server (BCS) \cite{BCS1,BCS2,BCS3}.  For the rest of the analysis we only need the irrep contents at the high symmetry points in the BZ. Adding the irreps given by  \textit{BandRep} for the two EBRs in Eq.~(\ref{MBRebrs}) yields 
\begin{eqnarray}\label{MIrreps}
 M(\Gamma)&=&\Gamma_{1}^+ (1)+\Gamma_{2}^- (1) +\Gamma_{3}^+ (2)+\Gamma_{3}^- (2)\nonumber\\
M(K)&=& 2K_3(2) + K_1(1) + K_2 (1) \\
M(M)&=&2M_1^+ (1)+ M_1^-(1) + M_2^+ (1)+ 2M_2^- (1),\nonumber
\end{eqnarray}
where the numbers in parenthesis give the dimensions of the irreps and lead to band crossings when greater than one. The irreps at the high symmetry lines and the compatibility relations are obtained by subduction from the high symmetry points and are also given by \textit{BandRep} \cite{BCS3}.

\begin{table*}[]
\centering
\begin{ruledtabular}
\begin{tabular}{cccc}
	Phase & Subset 1 & Subset 2 & Subset 3 \\ \hline
1 & $\Gamma_3^-+\Gamma_2^- ; K_2+K_3 ; M_1^+ + M_2^+ + M_2 ^- $ 
& $ \Gamma_1^+ ; K_1 ; M_1^+ $ 
& $\Gamma_3^+ ; K_3 ; M_1^- + M_2 ^- $ \\ \hline
	2 & $\Gamma_3^-+\Gamma_2^- ; K_2+K_3 ; M_1^- + 2M_2 ^- $ 
& $ \Gamma_1^+ ; K_1 ; M_1^+ $ 
& $\Gamma_3^+ ; K_3 ; M_1^+ + M_2 ^+ $ \\ \hline
	3 & $\Gamma_3^-+\Gamma_2^- ; K_2+K_3 ; M_1^+ + M_2^+ + M_2 ^- $ 
		& $\Gamma_3^+ ; K_3 ; M_1^- + M_2 ^- $ 
		& $ \Gamma_1^+ ; K_1 ; M_1^+ $ \\ \hline
	4 & $\Gamma_3^-+\Gamma_2^- ; K_2+K_3 ; M_1^- + 2M_2 ^- $ 
		& $\Gamma_3^+ ; K_3 ; M_1^+ + M_2 ^+ $ 
		& $ \Gamma_1^+ ; K_1 ; M_1^+ $ \\
\end{tabular}

\begin{tabular}{ccc}
	Phase & 	Subset 1 & Subset 2 \\ \hline
	5 & $\Gamma_3^-+\Gamma_2^- ; K_2+K_3 ; M_1^- + 2M_2 ^- $ 
		& $\Gamma_3^+ + \Gamma_1^+; K_3 + K_1 ; 2M_1^+ + M_2 ^+ $ \\ \hline
	6 & $\Gamma_3^-+\Gamma_2^- ; K_2+K_3 ; M_1^+ + M_2 ^+ +  M_2 ^- $ 
		& $\Gamma_3^+ + \Gamma_1^+; K_3 + K_1 ; M_1^+ + M_1^- + M_2 ^- $ \\ \hline
	7 & $\Gamma_3^-+\Gamma_2^- + \Gamma_3^+; K_2+2K_3 ; M_1^+ + M_1^- + M_2^+ + 2M_2 ^- $ 
		& $ \Gamma_1^+ ; K_1 ; M_1^+ $  \\ \hline
	8 & $\Gamma_3^-+\Gamma_2^- + \Gamma_1^+; K_1 + K_2+K_3 ; 2M_1^+ + M_2^+ + M_2 ^- $
		& $ \Gamma_3^+ ; K_3 ; M_1^- + M_2^- $ \\ \hline
	9 & $\Gamma_3^-+\Gamma_2^- + \Gamma_1^+; K_1 + K_2+K_3 ; M_1^+ + M_1^- + 2M_2 ^- $
		& $ \Gamma_3^+ ; K_3 ; M_1^+ + M_2^+ $ \\ \hline
	10 & $\Gamma_3^-+\Gamma_2^- + \Gamma_1^+; 2K_3 ; 2M_1^+ + M_2^+ + M_2 ^- $
		& $ \Gamma_3^+ ; K_1 + K_2 ; M_1^- + M_2^- $ \\ \hline
	11 & $\Gamma_3^-+\Gamma_2^- + \Gamma_1^+; 2K_3 ; M_1^+ + M_1^- + 2M_2 ^- $
		& $ \Gamma_3^+ ; K_1+K_2 ; M_1^+ + M_2^+ $ \\
\end{tabular}
\end{ruledtabular}
\caption{\label{table_irreps}Irreducible representations decomposition of the isolated subsets at the three high symmetry points in the BZ for all gapped phases. The subsets are ordered in terms of energy from lower to higher. The little groups for the $\Gamma$, K, M   points and T, T', $\Sigma$ lines are $D_{3d}$, $D_3$, $C_{2h}$ and $C_2$, $C_2$, $C_{1h}$, respectively.}
\end{table*} 

In order to find all the different gapped phases compatible with the symmetries, we just have to  order the irreps at the high-symmetry points in the BZ  in such a way that they lead to gaps in the phonon spectrum while respecting the compatibility relations arising from the subduction rules to the high symmetry lines. Moreover, an additional  constraint that distinguishes phonon spectra from electron bands is the existence of  three acoustic bands, for which  the dispersion relation must satisfy $\lim_{\textbf{k} \rightarrow 0} \omega(\textbf{k}) =0.$ The acoustic modes at  $\textbf{k}=0$ represent global  translations of the crystal and transform under the vector representation. Thus, by decomposing the vector representation $V_\Gamma$ of the little group of $\Gamma$, the irreducible representations of the acoustic bands at $\Gamma$ are obtained. In this case, $V_\Gamma$ decomposes as $V_\Gamma = \Gamma_2^- + \Gamma_3^- $ and we conclude that three eigenvalues associated with the irreps $\Gamma_2^- (1) $ and $\Gamma_3^- (2)$ must vanish at $\textbf{k}=0$. Table~\ref{table_irreps} gives all the irrep orderings that respect the compatibility relations and acoustic band constraints and result in gapped phases in the phonon spectrum.  It is important to note that we refer to these phases as ”gapped” because they contain isolated subsets of phonon bands separated by gaps,  notwithstanding the existence of gapless excitations in the form of acoustic modes.

\subsection{\label{TQC_analysis}Irrep-based topological analysis}

The next step is to find the isolated subsets of connected bands that can not  transform as  band representations.  As every band representation can be written as a sum of EBRs, if the irreps of an isolated subset can not be induced from any sum of EBRs then, according to TQC, the subset must have nontrivial topology. The result of this analysis is presented in Table \ref{table_EBRs}, where we see that phases $1,3,8,10$ and $11$ \textit{must} have 
nontrivial topology. As mentioned before, some of the  remaining seven  phases \textit{might} still be topologically nontrivial, but this cannot be diagnosed solely on the basis of their irrep contents. In the next subsection we will compute Wilson loops to diagnose their topology.

\begin{table*}
\begin{ruledtabular}
\begin{tabular}{cccc}
	Phase & Subset 1 & Subset 2 & Subset 3 \\ \hline
	{\bf 1} & $ B^- \mid_{3e} $ 
		& $ A_1^+ \mid_{1a} $ 
		& $A^+ \mid_{3e} - A_1^+ \mid_{1a} $ \\ \hline
	2 & $A_2^- \mid_{1a} +E^- \mid_{1a} $ 
		& $A_1^+ \mid_{1a} $ 
		& $E^- \mid_{1a} $ \\ \hline
      {\bf 3} & $ B^- \mid_{3e} $ 
		& $A^+ \mid_{3e} - A_1^+ \mid_{1a} $
		& $ A_1^+ \mid_{1a} $  \\ \hline
	4 & $A_2^- \mid_{1a} +E^- \mid_{1a} $ 
		& $E^+ \mid_{1a} $ 
		& $A_1^+ \mid_{1a} $ \\ \hline
	5 & $A_2^- \mid_{1a} +E^- \mid_{1a} $ 
		& $A_1^+ \mid_{1a}  + E^+ \mid_{1a} $
        & $ $ \\ \hline
	6 & $ B^- \mid_{3e} $ 
		& $A^+ \mid_{3e}$ 
        & $  $ \\ \hline
	7 & $A_2^- \mid_{1a}+ E^+ \mid_{1a} +E^- \mid_{1a} $ 
		& $A_1^+ \mid_{1a}$ 
        & $ $ \\ \hline
	 {\bf 8} & $A_1^+ \mid_{1a} + B^- \mid_{3e} $ 
		& $A^+ \mid_{3e} - A_1^+ \mid_{1a} $
        & $ $ \\ \hline
	9 & $A_1^+ \mid_{1a}+ A_2^- \mid_{1a} +E^- \mid_{1a} $ 
		& $E^+ \mid_{1a}$ 
        & $ $ \\ \hline
	 {\bf 10} & $A_1^+ \mid_{1a}+E_1^+ \mid_{1a}+E_1^- \mid_{1a}+A_1 \mid_{2d}-A^+ \mid_{3e} $ 
		& $ A_2^- \mid_{1a} - A_1 \mid_{2d} + A^+ \mid_{3e} $ 
        & $ $ \\ \hline
	 {\bf 11} & $ E^- \mid_{1a} +A_1 \mid_{2d} $ 
		& $A_1^+ \mid_{1a}+ A_2^- \mid_{1a}+ E^+ \mid_{1a} - A_1 \mid_{2d} $
        & $ $ \\
\end{tabular}
\end{ruledtabular}
\caption{\label{table_EBRs} Combinations of EBRs that reproduce the irrep content in Table \ref{table_irreps} for each of the isolated subsets, with the  Wyckoff positions indicated as subscripts. Note that these combinations are in general non-unique, but  have been presented as a sum whenever possible. Phases diagnosed as topological by TQC are given in boldface.}
\end{table*} 

The presence of negative coefficients in Table \ref{table_EBRs} for phases $1,3,8,10$ and $11$ is usually taken, at least for electronic bands, as a signature of fragile topology \cite{ wannier_obstructions, graphene_phonons}.
When the irreps of an isolated subset of bands can be obtained as a difference of EBRs, as in phase 1 in Table \ref{table_EBRs}, the addition of a trivial band that transforms under $A_1^+ \mid_{1a} $ in that case would ``trivialize" the fragile topology. For electrons this band may be found as core orbitals or  high energy conduction bands, but for phonons the number of bands is fixed and the required trivial band may not be available. This is another difference between electronic and phononic systems.

\section{\label{sec_analytic_model}
Dynamical matrix including third-nearest neighbor couplings}

\begin{figure*}[t!]
\includegraphics[width=0.9\linewidth]{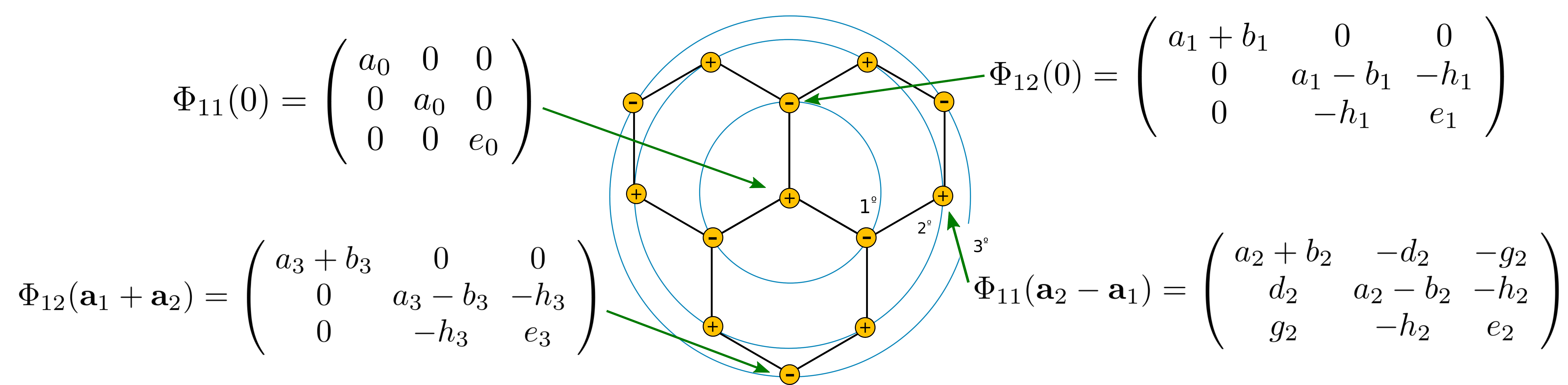}
\caption{\label{fig_neighbors} Matrices of force constants   consistent with the BHL symmetries up to third nearest neighbors. As described in Appendix \ref{app_apply_sym}, all the matrices for neighbors on the same  circle are related by symmetry operations to the one given in this figure.}
\end{figure*}

In this section we present a truncation of the dynamical matrix for the BHL that includes, up to third nearest neighbors,  all  couplings   compatible with the system symmetries. We stop at third nearest neighbors because that is enough to reach all the phases in Tables~I and~II. The truncated dynamical matrix provides an analytical model that  is used below   to compute Wilson loops \cite{wilson1,wilson2,wilson3} for  every disconnected subset of bands,  identifying thus all the topologically non-trivial phases.

The harmonic potential energy of the crystal can be written as
\begin{equation}
U^h=\frac{1}{2} \sum_{\textbf{RR}',ij} \textbf{u}_i(\textbf{R}) \Phi_{ij}(\textbf{R}-\textbf{R}')\textbf{u}_j(\textbf{R}'),
\end{equation}
where $\textbf{R}$ and $\textbf{R}'$ label two distinct unit cells, $\textbf{u}$ is a displacement vector, $(i,j)$ are the atom indices and $\Phi_{ij}$ is a 3$\times$3 matrix of force constants which  must be real by time reversal symmetry. Note that the fact that $U^h$ is a quadratic form in the atomic displacements implies 
\begin{equation}
\Phi_{ij}(\textbf{R}) = \Phi_{ji}^t(-\textbf{R}), \label{sym1}
\end{equation}
where $t$ indicates matrix transposition. As shown in Appendix \ref{app_apply_sym}, after restricting ourselves to third nearest neighbors and applying all the symmetry constraints, we are left with sixteen independent parameters:
$$ (\overbrace{a_0,e_0},\overbrace{a_1,b_1,\underline{h_1},e_1},\overbrace{a_2,b_2,e_2,\underline{g_2},d_2,\underline{h_2}},\overbrace{a_3,b_3,e_3,\underline{h_3}}) ,$$
where a subindex $n$ indicates a coupling between $n$th-nearest neighbors
and the underlined parameters describe  couplings between on- and off-plane phonons that vanish  for the planar honeycomb lattice \cite{graphene_phonons}. The corresponding matrices of coupling constants are shown in  Fig \ref{fig_neighbors}.
As we will see, not all the parameters are independent due to the additional constraints imposed by the existence of  three acoustic bands.

The dynamical matrix is defined as a Fourier transform in the usual way
\begin{equation}\label{Dyn_matrix}
D_{ij}(\textbf{k}) = \sum_{\textbf{R}} \frac{\Phi_{ij}(\textbf{R})}{\sqrt{M_i M_j}}e^{-i \textbf{k} \cdot \textbf{R}},
\end{equation}
where \textbf{k} belongs to the first Brillouin zone and $M_i$ is the mass of atom $i$. After analytically diagonalizing the dynamical matrix at the high symmetry points as shown in Appendix \ref{app_find_eigen}, we take care of  the existence of  acoustic branches by imposing $w^2(\Gamma_3^-)=w^2(\Gamma_2^-)=0$, which will be satisfied as long as
\begin{equation}
\begin{aligned}
& a_0 = -3(a_1+2a_2+a_3) ,\\
& e_0 = -3(e_1+2e_2+e_3).
\end{aligned}
\end{equation}
This leaves  14 independent parameters that  can be  tuned to replicate any of the eleven gapped phases or fitted to  experimental or DFPT data for real materials:
\begin{equation}\label{forteen}
 (\overbrace{a_1,b_1,\underline{h_1},e_1},\overbrace{a_2,b_2,e_2,\underline{g_2},d_2,\underline{h_2}},\overbrace{a_3,b_3,e_3,\underline{h_3}}) 
 \end{equation}

We close the discussion of the $6\times 6$ dynamical  matrix $D(\textbf{k})$ for the BHL by noting that, in general, we can not  expect  to diagonalize it analytically, as  that requires the solution of a sixth order polynomial equation. However, while this is true for generic points in the BZ, it is actually possible to obtain explicit expressions for all  the frequencies and eigenmodes at the three high symmetry points in the BZ. This is a simple consequence of Wigner's theorem~\cite{bradley_cracknell}, which applied to  the dynamical matrix  stablishes that changing to a basis of symmetry adapted modes reduces $D(\textbf{k})$ to a block-diagonal form. Specifically, each irrep of dimension $d$ and multiplicity $m$ gives rise to $d$ identical $m\times m$ blocks in $D(\textbf{k})$. A look at Eq.~(\ref{MIrreps}) shows that the largest multiplicity is two, which involves solving at most a quadratic equation. The process of diagonalization using Wigner's theorem requires the construction of symmetry-adapted modes, which are given in Appendix~\ref{app_find_eigen} together with the resulting analytic formulas for the frequencies. As according to Table~\ref{table_irreps}   the topology of phononic bands is largely determined by the ordering of frequencies at the high symmetry points, having explicit formulas  greatly simplifies the study of  the phase space.

\subsection{\label{sec_wilson}Wilson loop windings}

\begin{figure}[]
\includegraphics[width=8.5cm,keepaspectratio]{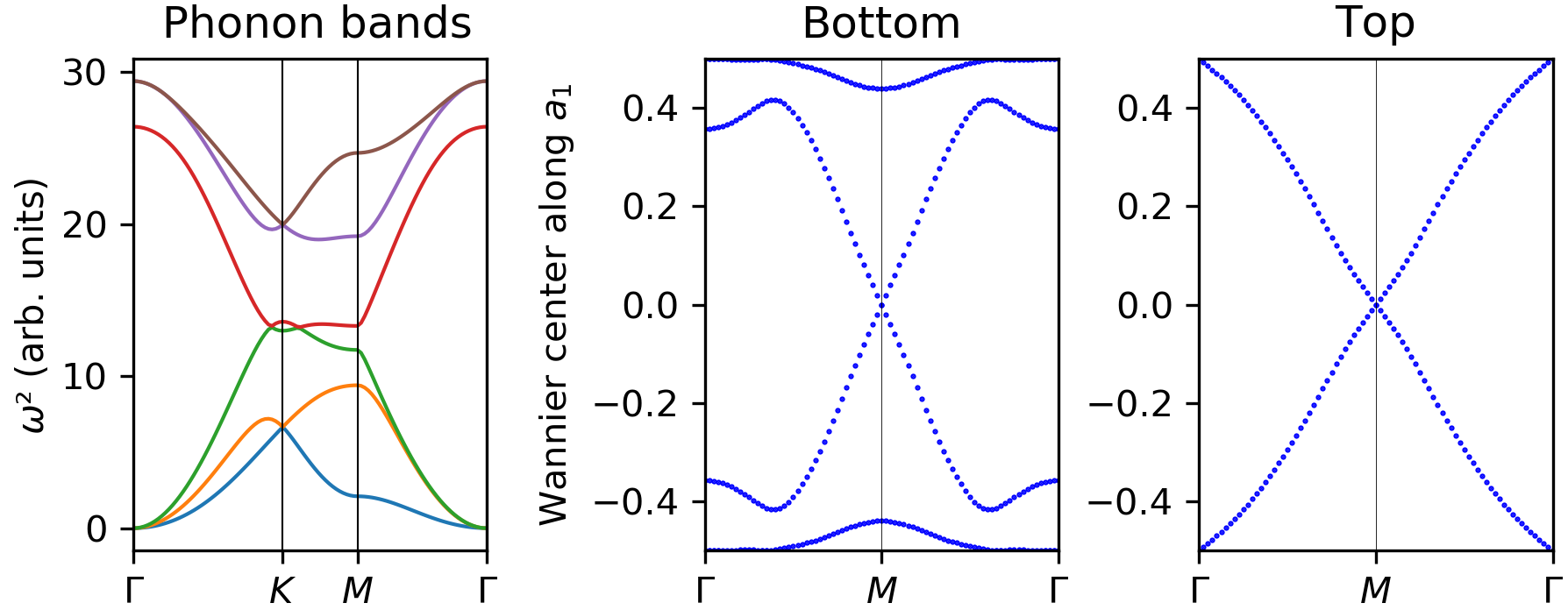}
\caption{\label{fig_Wilson} Wilson loop analysis for phase 8 in Tables \ref{table_irreps} and \ref{table_EBRs}. Phonon bands (left) and Wannier centers for the bottom (middle) and top (right) subsets of bands. The WL spectrum of the bottom bands shows trivial topology, while being non-trivial (winding) at the top subset.}
\end{figure}

Another benefit of having an analytical model is that Wilson loops (WL) can be easily computed using a tigh binding code such as PythTB \cite{PythTB}. The existence of windings in the WL spectrum  that cannot be eliminated by any perturbation that respects  the  symmetries of the system and does not close a gap guarantees that the subset of bands has nontrivial topology.
We will consider a \hbox{$\mathbf{b_1}$-oriented} Wilson loop~\cite{wilson1,wilson2} defined by
\begin{equation}
W(k_2)=P \ e^{-\int_{0}^{2\pi}dk_1 A_{i,j}(\mathbf{k}) },
\end{equation}
where $P$ means that the integral is path-ordered and $A_{i,j}(\mathbf{k})=\braket{u_i(\mathbf{k})|\partial_{k_1}|u_j(\mathbf{k})}$ is the non abelian Berry connection built from the normal modes $u_i(\mathbf{k})$ of a subset of isolated bands. The eigenvalues of this WL matrix are of the form $e^{i2\pi x_1(k_2)}$, where $x_1(k_2)$ are the positions of the hybrid Wannier functions \cite{taherinejad_2014}
along $\mathbf{a_1}$. As   $k_2$ moves along a clossed path ($\Gamma$-M-$\Gamma$), these Wannier centers move along the $\mathbf{a_1}$ direction as shown in Fig \ref{fig_Wilson}. 

According to the results in Table \ref{table_EBRs}, most of the subsets of bands \textit{might} transform as a band representation and therefore the corresponding phases could   have trivial topology. However, after realizing all the phases within the model, we were able to compute the Wilson loop spectrum for all the subsets as shown in Appendix \ref{app_WL}. The results imply that nine out  eleven gapped phases have subsets of bands with winding in the WL spectrum  and  are thus topologically nontrivial, as shown in Table \ref{table_wilson}. Notice also that all  the phases predicted to be topological with TQC techniques in Table \ref{table_EBRs} have indeed non-zero windings in the WL spectra.

\begin{table}[h]
\begin{ruledtabular}
\begin{tabular}{cccccccccccc}
	\textbf{Phase} & 				\textbf{1} & 	\textbf{2} & \textbf{3 }&	 \textbf{4}& 	\textbf{5}& \textbf{6}& 	\textbf{7}&	 \textbf{8}& \textbf{9}& \textbf{10}& \textbf{11} \\ \hline
   	Subset 1 	&0 		& 0 	& 0 	& 2		& 0		& 0		& 0& 	0& 	0& 	0& 2 \\
    Subset 2 	& 0 	& 0 	& 1 	& 2		& 2 	& 0		& 0&	 1&	 2& 1& 2 \\
	Subset 3 		  	& 1 	& 2 	& 0 	& 0	 	& $- $ & $ -$&  $-$&  $ -$&  $ -$&  $ -$&  $ -$ \\
\end{tabular}
\end{ruledtabular}
\caption{\label{table_wilson}Windings in the WL spectrum of the  isolated subsets of bands in the eleven phases. All phases with non-zero winding are topological. }
\end{table}

\section{\label{sec_DFPT}Merging the ab initio and analytical model results}

\subsection{\label{DFPT_results}DFPT results}

\begin{figure*}[t!]
\includegraphics[width=0.9\linewidth]{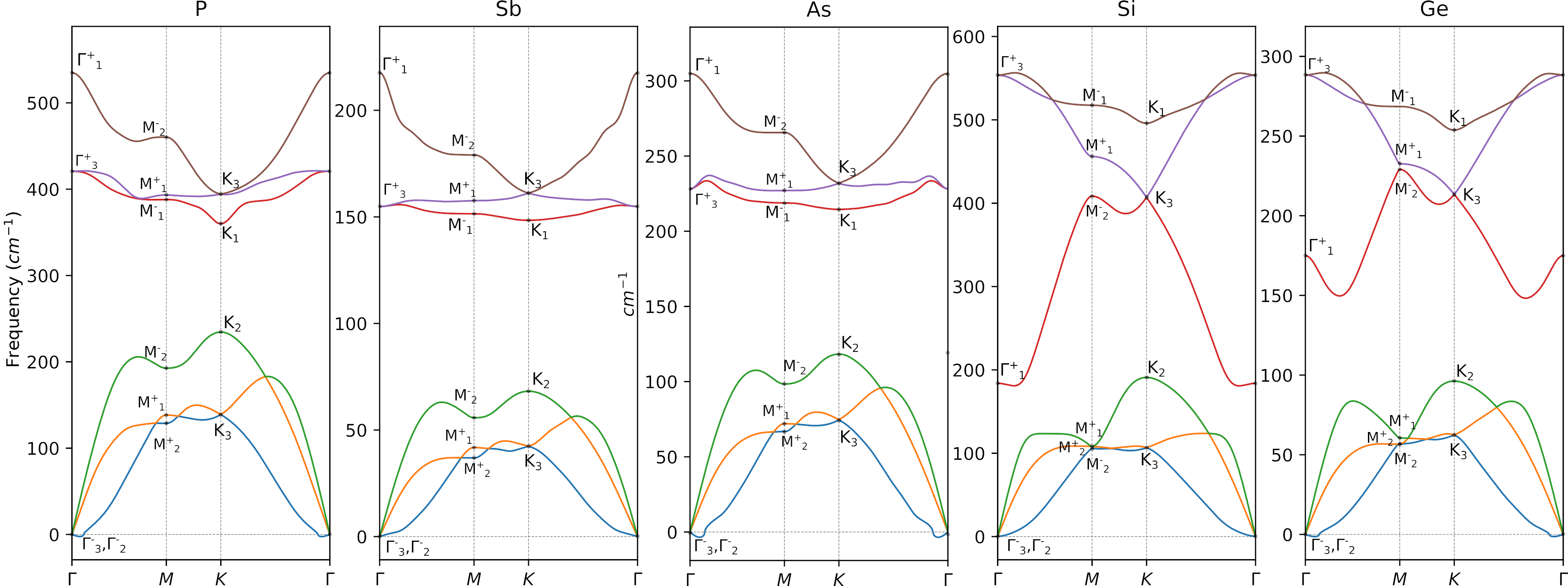}
\caption{\label{fig_phonons} DFPT phonon dispersion bands of P, Sb, As, Si and Ge in the BHL. The irrep content of each subset of bands corresponds with phase 6 in Table \ref{table_irreps} for all cases, which is not topological according to the results in Table \ref{table_wilson}.
}
\end{figure*}

In this section we relate our previously developed model to  the phonon spectra of real materials with the buckled honeycomb lattice. To search for phononic topological phases we use a procedure based on the following steps: (i) relax the structure, (ii) compute the phonon spectrum with DFPT and check whether it is gapped, (iii) place the material on the phase diagram by comparing the irreps of the computed dispersion bands with the ones in Table \ref{table_irreps}.

\begin{figure}[]
\includegraphics[width=8.6cm,keepaspectratio]{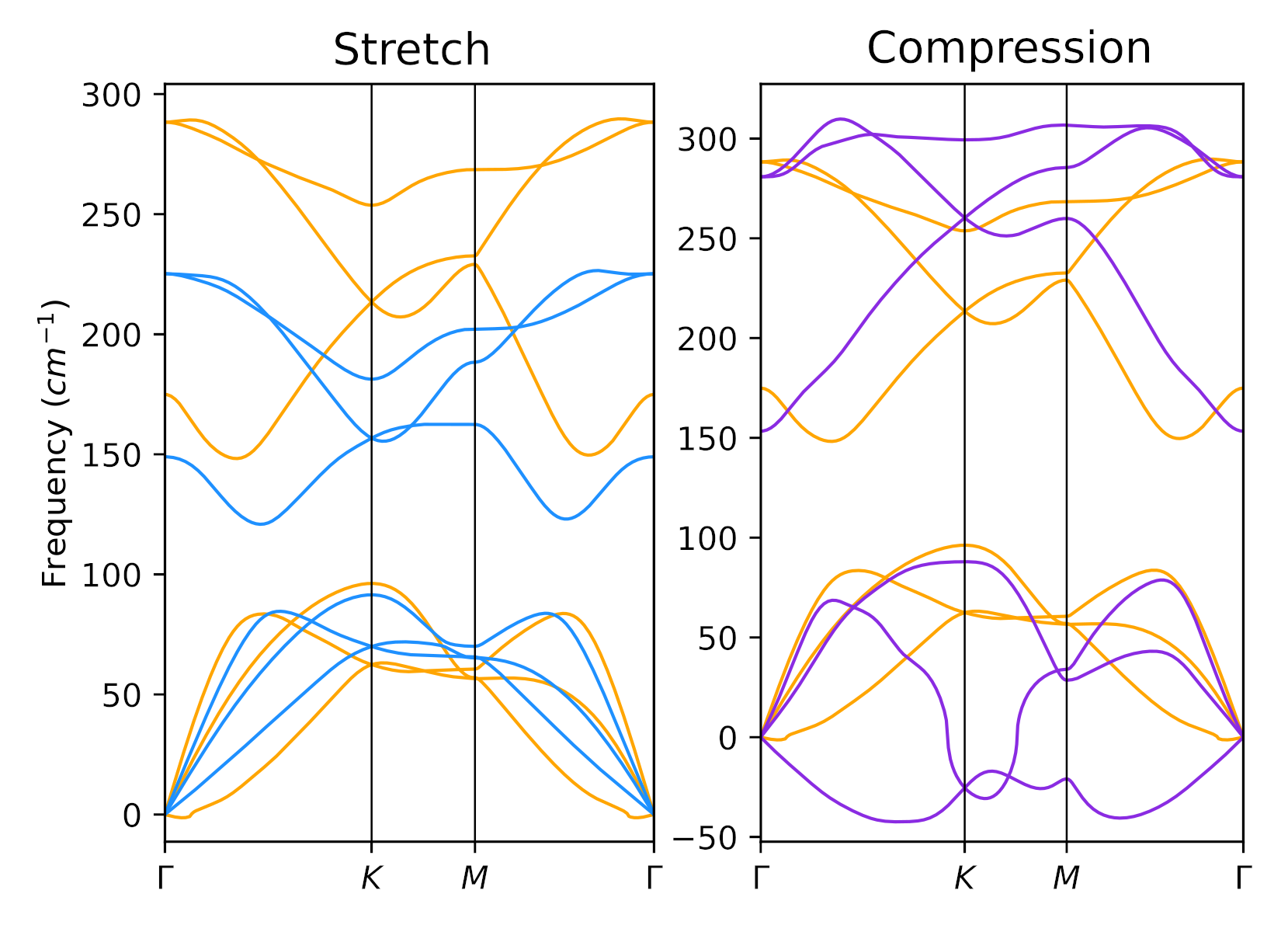}
\caption{\label{fig_strain} Phonon dispersion bands for Germanium BHL (orange), with a 7 \% stretch (blue) and 7 \% compression (purple).
}
\end{figure}

We searched for topological behaviour in Si, Ge, P, As, and Sb. The calculated phonon spectra and irreps as provided by  the {\sc Quantum ESPRESSO} package \cite{giannozzi2009quantum,QE-2017} are displayed in Fig \ref{fig_phonons}, and a comparison with Table \ref{table_irreps} shows that all of them are in phase 6, which is topologically  trivial. However, given that nine out of eleven possible  phases are topologically nontrivial, a natural question is whether a  transition to a topological phase could be induced  by some kind of symmetry preserving method such  as isotropic strain,  doping and photoexcitation. To test this possibility we numerically simulated 
stretching or compressing the lattice by up to a 7\%, which is already a rather large deformation for experimental setups \cite{Peng2020Strain}, but found that the materials remained in the topologically trivial phase 6. Indeed, as seen for example in Fig.~\ref{fig_strain}  for Germanium, even such large deformations are unable to close any of the gaps at the HSPs and cause a band inversion, which would be necessary in order to change the band topology.
This is due to the fact that all the gaps at the HSPs are a large fraction of the total span of the bands, so that any band inversion would require  large relative changes in the frequencies, which are very hard to achieve  experimentally. In other words, all five materials are physically very far from any topologically nontrivial phase. %

\subsection{Monte Carlo analysis}

In order to understand why all the materials studied by DFPT are so far from any topologically nontrivial phase
we carried out a Monte Carlo study of the phase space of the analytical model. This is possible because we have analytic expressions for the frequencies at the HSPs as functions of the force constants, as given explicitly  in Appendix~\ref{app_find_eigen}. Thus, to any point in the phase space specified by the values of the forteen independent
coupling constants $(a_1,b_1,....,h_3)$ in Eq.~(\ref{forteen}) we can associate another point $(w_{\Gamma_3^+},w_{\Gamma_3^+},w_{K_{3,1}},w_{K_{3,2}},...,w_{M_2^-})$ in the space of nonzero frequencies at the HSPs. Note that we are discarding the acoustic phonon  frequencies $(w_{\Gamma_3^-},w_{\Gamma_2^-})$, as they  must vanish  at the $\Gamma$ point.

It is important to note that the map from the fourteen independent parameters to the  twelve nonzero frequencies at the HSPs is \textit{not} invertible. Thus, there is no direct way to find a model specified by a given set of frequencies. Instead, we must do a Monte Carlo sampling of parameter space, compute the associated frequencies  and select the point that most closely satisfies our requirements. This is precisely the method we followed to find model realizations for the eleven gapped phases in Table~\ref{table_irreps}, with the resulting phonon bands and Wilson loops given in Appendix~\ref{app_WL}. 
Comparing the computed frequencies with the irreps in Table~\ref{table_irreps} would immediately assign a phase to the model.

\begin{figure}[t]
\includegraphics[width=8.5cm,keepaspectratio]{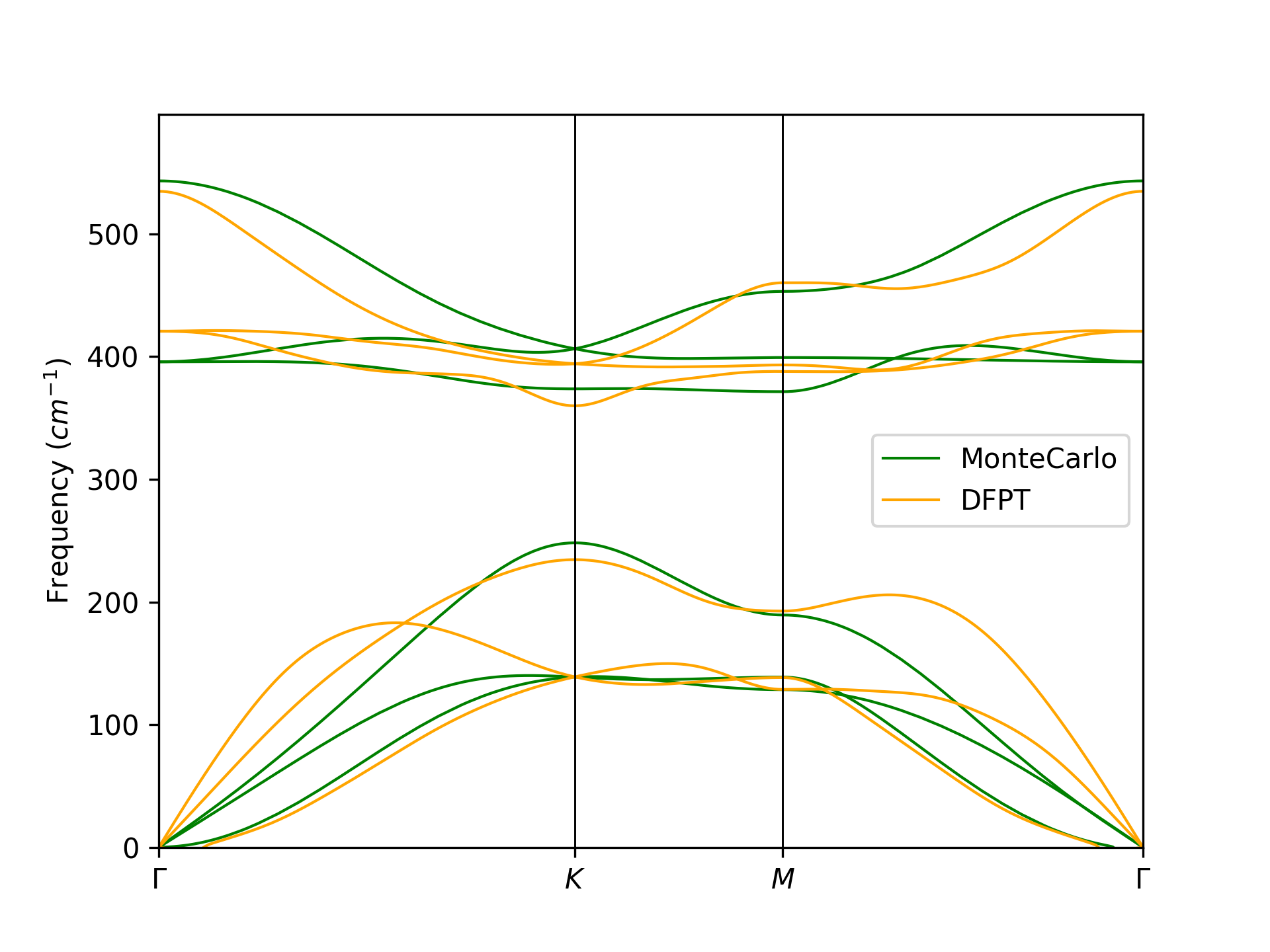}
\caption{\label{fig_mode_fit} DFPT dispersion phonons for the ``blue phosphorus" structure compared with the dispersion bands of a model with $r=0.054$.}
\end{figure}

The method can also be used to find the model that best fits a real material. First, we define the following distance in the space of frequencies

\begin{equation}\label{error}
 r \equiv \frac{1}{12} \sum_{i=1}^{12} \frac{|w^2_i-\lambda^2_i|}{w_i\lambda_i},
\end{equation}
where $w_i$ are the frequencies   at the HSPs for the real material, $\lambda_i$  the frequencies for a model obtained by Monte Carlo sampling, and $i$ labels the twelve nonzero frequencies at the HSPs.
Therefore, this distance $r$ measures the overall relative difference between the frequencies of the real material and a particular model. Once a model with sufficiently small $r$ is obtained by random sampling of parameter space, we can use the gradient method to minimize the value of $r$ and improve the fit. For example, the best fit ($r =0.054$) for phosphorous is shown in 
Fig.~\ref{fig_mode_fit}.

Finally, a relation can be established between the existence of  topological phases and the need for sizeable couplings between far apart  neighbors. In any real material the couplings between different neighbors are bound to decrease as the distance is increased, due to the localized nature of the atomic wave functions. In the case of the planar honeycomb lattice, where the boundaries between different phases can be given analically, it was found that the existence of two out of  four possible topological phases required  a relatively large third 
nearest neighbor force constant~\cite{graphene_phonons}.

By fitting our DFPT results to the analytical model, as explained above, we find that in all cases  the force constants decay faster than according to a (100:35:4) ratio for onsite, first and second neighbors respectively. We have explored the model parameter space using different decay ratios as thresholds (see Appendix \ref{app_top_neigh} for details) and computed the percentage of space occupied by each  phase. The results in Table \ref{table_phase_space} show that some of the predicted phases may be almost impossible to realize for force constants restricted to decay as in real materials, although this does not exclude the possibility of metamaterial realizations. Moreover, it shows that as we approach realistic decay ratios (third column), the volume occupied by the trivial phase 6 increases. This is is consistent with the fact that  all the  materials considered in this paper happen to be in phase~6.

We close this analysis by noting that the phases that are  most suppressed  for fast decay ratios (column three in Table \ref{table_phase_space}) are precisely the ones with higher WL windings in table \ref{table_wilson}. This correlation admits a simple heuristic explanation. The dynamical matrix $D_{ij}(\textbf{k})$ is periodic in reciprocal space and therefore admits a Fourier series representation, with force constants between far away neighbors contributing to higher harmonics. On the other hand, WL windings measure the extend to which the eigenvectors of the dynamical matrix twist around as we move across de BZ, 
and this twisting is obviously related to the harmonics in the in $D_{ij}(\textbf{k})$. Thus, high WL windings are favored by 
strong force constants between far away neighbors.

\begin{table}
\begin{ruledtabular}
\begin{tabular}{cccc}
 	\textbf{Ratio} & \textbf{100:50:10} 	& \textbf{	100:40:7} 	& 	\textbf{100:35:4} \\ \hline
	Phase 1 &  1.11\%	& 1.361\% 	& 4.81\% 	\\
	Phase 2 & 0.06\% 	& 0.05\% 	& 0\% \\
	Phase 3 & 9.71\% 	& 11.9\% 	& 3.09\% 	\\
	Phase 4 & 0.15\%  & 0.14\%  & 0\%	 \\
	Phase 5& 0.19\%	& 0.17\%  & 0\% \\
	Phase 6& 30.03\%	& 49.85\% & 89.05\%	\\
	Phase 7& 10.14\%   & 5.88\%  & 0.06\%  \\
    Phase 8& 1.89\%    & 1.22\%  & 2.71\%  \\
	Phase 9& 19.87\%   & 15.06\% & 0.03\% \\
	Phase 10& 1.47\%   & 0.43\%  & 0.73\% \\
	Phase 11& 25.31\%   & 13.89\% & 0.02\% \\
\end{tabular}
\end{ruledtabular}
\caption{\label{table_phase_space}Percentage of randomly sampled points belonging to each  gapped phase. For each column the force constants are constrained to decay faster than the indicated decay ratio, which  compares the   mean   absolute values of  non-zero force constants for onsite, first and second neighbors. The mean value of third nearest neighbor constants is constrained to be smaller than that of  second neighbors.} %
\end{table}

\section{\label{sec_conclusions}Conclusions}

We have predicted all the possible topological phases for phonons on the buckled honeycomb lattice. To this end we have used TQC group theory techniques and discussed how they may be applied to phonons. Eleven gapped phases where found with nine of them being topological. This result, together with the ones for the planar honeycomb lattice \cite{graphene_phonons}, suggests that a huge array of topological phases exist for distinct structures. This has been confirmed in the recently launched catalogue of 3D materials with topological phonons \cite{phonon_catalogue}.

Finally, we have constructed the most general dynamical matrix  for the buckled honeycomb lattice including all couplings compatible with the symmetries up to third nearest neighbors. The model has been used to fully characterize the  topology of the possible phases using Wilson loops and to analyze the complete phase space under conditions resembling real materials. We have studied the possibility of having topological phononic phases in Si, Ge, P, As, and
Sb in the buckled honeycomb structure and explained why  inducing topological phases in these systems is difficult. 

\begin{acknowledgments}
 M.G.V. and J.L.M. thank B.A. Bernevig, L. Elcoro and Zhida Song for helpful discussions. M.G.V., I. E. and M.G.A acknowledge the Spanish Ministerio de Ciencia e Innovacion (grant PID2019-109905GB-C21). M.G.V. is also thankful to the Deutsche Forschungsgemeinschaft (DFG, German Research Foundation) GA 3314/1-1 – FOR 5249 (QUAST). The work of J.L.M. has been supported by Spanish Science Ministry grants PGC2018-094626-B-C21 and PID2021-123703NB-C21 (MCIU/AEI/FEDER, EU), and Basque Government grants IT979-16 and IT1628-22. This work is part of a project that has received funding from the European Research Council (ERC) under the European Union’s Horizon 2020 research and innovation program (grant agreement no. 101020833)
\end{acknowledgments}

\appendix

\section{\label{app_methods}Methods}
All  density functional-perturbation theory (DFPT) \cite{RevModPhys.73.515} calculations were done using the {\sc Quantum ESPRESSO} package \cite{giannozzi2009quantum,QE-2017}. We parametrize the exchange-correlation functional assuming the Perdew-Burke-Ernzerhof \cite{PBE_exchange} parametrization and model the electron-ion interaction with projector augmented wave pseudopotentials \cite{PAW_Kresse1,PAW_Kresse2} including four electrons in the valence for Si and Ge, and three electrons for P, As, and Sb. We use a kinetic energy cutoff of 60 Ry for the plane-wave basis and 600 Ry for the charge density. Brillouin zone integrals in the DFPT self-consistent loop were calculated with a $20\times 20\times 1$ grid and the occupancies have a Methfessel-Paxton first-order spreading \cite{mp-smearing-QE} of 0.02 Ry. Prior to the phonon calculation, the structures were relaxed to the Born-Oppenheimer minimum. Then we calculated the the force constants in a $12\times 12\times 1$ grid and obtained the phonon spectra by Fourier interpolation.

\section{\label{app_apply_sym} Symmetry constraints on the analytical model}

In this section we compute the four matrices of force constants in Fig.~\ref{fig_neighbors} and show how to use them to obtain the matrices for symmetry-related neighbors. Then the Fourier transform in Eq.~(\ref{Dyn_matrix}) yields the most general dynamical matrix $D(\textbf{k})$ compatible with all the symmetries and including up to third nearest neighbor couplings.
A generic  force constants matrix is parametrized by nine real constants
\begin{equation}
\Phi_{ij}(\textbf{R})= \left( \begin{array}{ccc}
	a +b & -c-d & -f-g \\
	-c+d & a -b & -h-i \\
	-f+g & -h+i & e
\end{array} \right),
\end{equation}
where the indices $i$ and $j$ take the values $1$ or $2$ for the two atoms in the unit cell (see Fig.~\ref{fig_vectors}), and $\textbf{R}$ is a lattice vector connecting the origins of the cells to which the two atoms belong. 

\begin{figure}[b]
\includegraphics[width=8.5cm,keepaspectratio]{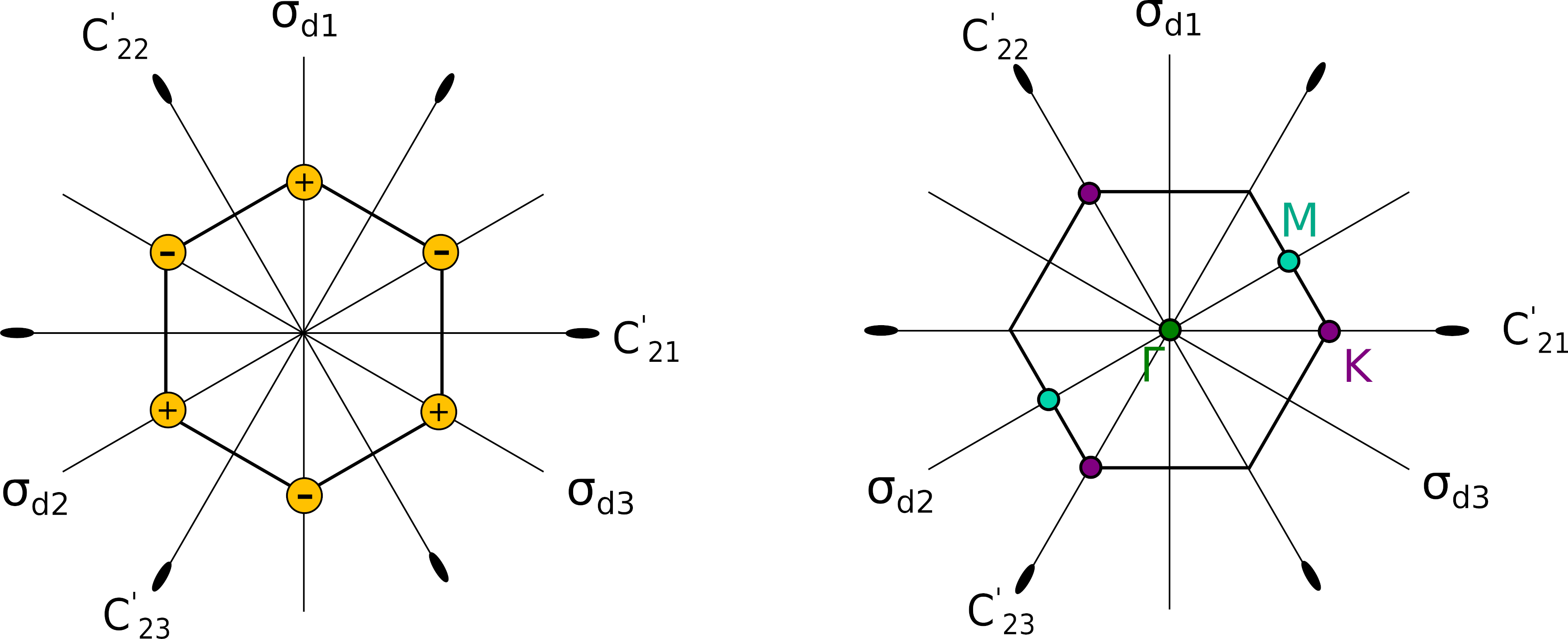}
\caption{\label{fig_sym_operations} Symmetry operations of the $D_{3d}$ group in real and reciprocal space. The threefold rotation axes $C_{3}^\pm$ are perpendicular to the plane of the crystal.}
\end{figure}

Besides the constraint in Eq.~(\ref{sym1}), the force constants matrices must satisfy additional relations  due to the point group symmetries of the crystal~\cite{RevModPhys.40.1}
\begin{equation}
\Phi_{i'j'}(S\textbf{R}) = V(S)\Phi_{ij}(\textbf{R})V(S)^t, \label{sym2} 
\end{equation}
where $V(S)$ is the $3\times 3$ matrix for the symmetry operation $S$ in the vector representation; $i',j'$ are the indices for the atoms in the transformed positions; and $S\textbf{R}$ is the lattice vector connecting the origins of the cells to which the atoms in the transformed positions belong. Note that, in general, $S\textbf{R}\neq V(S)\textbf{R}$.

It is important to note that, given that the space group for the BHL is symmorphic,  we can restrict ourselves to point group operations. This would not be possible for a nonsymmorphic group, where the point group is \textit{not} a subgroup of the space group.

\subsection{On-site couplings}

According to Eq.~(\ref{sym2}) the matrix $\Phi_{11}(\textbf{0})$ must be invariant under any operation belonging to the site symmetry group $C_{3v}$ of atom $1$,
\begin{equation}
\Phi_{11}(\textbf{0}) = V(S)\Phi_{11}(\textbf{0})V(S)^t\;\;  \forall\; S\in C_{3v}.
\end{equation}
Applying this equation with  $S$ equal to $C_3^+$ and $\sigma_{d_1}$ (see Fig.~(\ref{fig_sym_operations}) for notation), which together generate the group $C_{3v}$, shows that the on-site matrix must take the form
\begin{equation}
\Phi_{11}(0) = \left( \begin{array}{ccc}
	a_0 & 0 & 0 \\
	0 & a_0 & 0 \\
	0 & 0 & e_0
\end{array} \right).
\end{equation}
Noting that the spatial inversion $I$ exchanges atoms $1$ and $2$  and using Eq.~(\ref{sym2}) for $S\!=\!I$ yields
\begin{equation}
\Phi_{22}(\textbf{0}) = V(I)\Phi_{11}(\textbf{0})V(I)^t = \Phi_{11}(\textbf{0}). 
\end{equation}

\subsection{First nearest neighbors}

Here we consider the couplings between atoms $1$ and $2$ in Fig.~\ref{fig_vectors}. 
We can obtain a first constraint on $\Phi_{11}(\textbf{0})$ by using Eq.~(\ref{sym2}) with $S=\sigma_{d_1}$, which leaves atoms $1$ and $2$ invariant
\begin{equation}\label{first1}
\Phi_{12}(\textbf{0}) = V(\sigma_{d_1})\Phi_{12}(\textbf{0})V(\sigma_{d_1})^t .
\end{equation}
A second constraint is obtained by combining Eq.~(\ref{sym2}) for $S=I$ with Eq.~(\ref{sym1})
\begin{equation}\label{first2}
\Phi_{12}(\textbf{0}) = V(I)\Phi_{21}(\textbf{0})V(I)^t =\Phi_{21}(\textbf{0})=\Phi_{12}(\textbf{0})^t,
\end{equation}
and the two constraints together  imply
\begin{equation}\label{Phi12nn}
\Phi_{12}(\textbf{0}) = \left( \begin{array}{ccc}
	a_1+b_1 & 0 & 0 \\
	0 & a_1-b_1 & -h_1 \\
	0 & -h_1 & e_1
\end{array} \right).
\end{equation}
Finally, using  Eqs.~(\ref{sym1}) and (\ref{sym2}) for $S=C_3^\pm$  yields the remaining nearest neighbor matrices
in terms of  Eq.~(\ref{Phi12nn})
\begin{equation}
\begin{aligned}
 \Phi_{12}(\textbf{a}_1)&=V(C_3^+) \Phi_{12}(\textbf{0}) 	V(C_3^+)^t \\
 \Phi_{12}(\textbf{a}_2)&=V(C_3^-) \Phi_{12}(\textbf{0}) 	V(C_3^-)^t \\
 \Phi_{21}(\textbf{0}) &=\Phi_{12}(\textbf{0})  \\
 \Phi_{21}(-\textbf{a}_1)&= \Phi^t_{12}(\textbf{a}_1) \\
 \Phi_{21}(-\textbf{a}_2)&=\Phi^t_{12}(\textbf{a}_2) .
\end{aligned}
\end{equation}
\subsection{Second nearest neighbors}

Using Eq.~(\ref{sym2}) with $S=\sigma_{d_1}$, that exchanges two second neighbors, followed by Eq.~(\ref{sym1}), gives
\begin{equation}
\sigma_{d_1} \Phi_{11}(\textbf{a}_2-\textbf{a}_1) \sigma^{-1}_{d_1}=\Phi_{11}(\textbf{a}_1-\textbf{a}_2)=\Phi^t_{11}(\textbf{a}_2-\textbf{a}_1),
\end{equation}
which directly leads to:
\begin{equation}
\Phi_{11}(\textbf{a}_2-\textbf{a}_1) = \left( \begin{array}{ccc}
	a_2+b_2 & -d_2 & -g_2 \\
	d_2 & a_2-b_2 & -h_2 \\
	g_2 & -h_2 & e_2
\end{array} \right).
\end{equation}
Then the remaing second nearest neighbor matrices can be obtained using    $\Phi_{22}(\textbf{a}_1-\textbf{a}_2)=  \Phi_{11}(\textbf{a}_2-\textbf{a}_1)$ and
\begin{equation}
\begin{aligned}
& \Phi_{11}(-\textbf{a}_2)= \Phi_{22}(\textbf{a}_2)= V(C_3^+) \Phi_{11}(\textbf{a}_2-\textbf{a}_1) V(C_3^+)^t	\\
& \Phi_{11}(\textbf{a}_1)=\Phi_{22}(-\textbf{a}_1)= V(C_3^-) \Phi_{11}(\textbf{a}_2-\textbf{a}_1) V(C_3^-)^t	\\
& \Phi_{11}(\textbf{a}_1-\textbf{a}_2)= \Phi_{22}(\textbf{a}_2-\textbf{a}_1)=  \Phi^t_{11}(\textbf{a}_2-\textbf{a}_1)	\\
& \Phi_{11}(\textbf{a}_2)= \Phi_{22}(-\textbf{a}_2)=  \Phi^t_{11}(-\textbf{a}_2)	\\
& \Phi_{11}(-\textbf{a}_1)=\Phi_{22}(\textbf{a}_1)=  \Phi^t_{11}(\textbf{a}_1).	\\
\end{aligned}
\end{equation}

\subsection{Third nearest neighbors}
As seen in Fig.~\ref{fig_neighbors}, the geometry of third nearest neighbors is closely related to the one for first neighbors, with $\sigma_{d_1}$ and the inversion $I$ playing analogous roles here. Instead of Eqs.~(\ref{first1}) and (\ref{first2}) we have now
\begin{equation}
\Phi_{12}(\textbf{a}_1+\textbf{a}_2)= V(\sigma_{d_1})\Phi_{12}(\textbf{a}_1+\textbf{a}_2)V(\sigma_{d_1})^t 
\end{equation}
and 
\begin{equation}
\Phi_{12}(\textbf{a}_1+\textbf{a}_2) = \Phi_{21}(-\textbf{a}_1-\textbf{a}_2) = \Phi^t_{12}(\textbf{a}_1+\textbf{a}_2),
\end{equation}
and these two conditions imply
\begin{equation}
\Phi_{12}(\textbf{a}_1+\textbf{a}_2) = \left( \begin{array}{ccc}
	a_3+b_3 & 0 & 0 \\
	0 & a_3-b_3 & -h_3 \\
	0 & -h_3 & e_3
\end{array} \right).
\end{equation}
The remaing third nearest neighbor matrices are given by
\begin{equation}
\begin{aligned}
 \Phi_{12}(\textbf{a}_2-\textbf{a}_1)&=V(C_3^+) \Phi_{12}(\textbf{a}_1+\textbf{a}_2)	V(C_3^+)^t \\
 \Phi_{12}(\textbf{a}_1-\textbf{a}_2)&=V(C_3^-) \Phi_{12}(\textbf{a}_1+\textbf{a}_2)	V(C_3^-)^t \\
 \Phi_{21}(-\textbf{a}_1-\textbf{a}_2)&= \Phi^t_{12}(\textbf{a}_1+\textbf{a}_2) \\
 \Phi_{21}(\textbf{a}_1-\textbf{a}_2)&= \Phi^t_{12}(\textbf{a}_2-\textbf{a}_1) \\
 \Phi_{21}(\textbf{a}_2-\textbf{a}_1)&=\Phi^t_{12}(\textbf{a}_1-\textbf{a}_2). 
\end{aligned}
\end{equation}

We finish by giving the matrices of the vector representation used in this Appendix
\begin{equation}
V(\sigma_{d_1})\! =\! \left( \begin{array}{ccc}
	-1 & 0 & 0 \\
	0 & 1 & 0 \\
	0 & 0 & 1
\end{array} \right)\!, 
V(C_3^+)\! =\! \left( \begin{array}{ccc}
	-\frac{1}{2} & -\frac{\sqrt{3}}{2} & 0 \\
	\frac{\sqrt{3}}{2} & -\frac{1}{2} & 0 \\
	0 & 0 & 1
	\end{array} \right).
\end{equation}
Note also  that $V(C_3^-)=V(C_3^+)^t$ and $V(I)=-\openone_3$.

\section{\label{app_find_eigen}Spectrum of the dynamical matrix at the high symmetry points of the BZ}

As reviewed in Section~\ref{sec_analytic_model}, group theory can be used to simplify the diagonalization of the dynamical matrix by expressing it in a basis of symmetry-adapted modes, where it takes a block-diagonal form.

\subsection{Symmetry adapted modes}

The mechanical representation is induced from the vector representation of $C_{3v}$, which is the site-symmetry group for the WP $2b$. The vector representation is reducible, and according to the BCS
\begin{equation}
V=A_1(z) + E(x,y).
\end{equation} 
As a consequence, we may compute separately the off-plane modes, induced from $A_1$ and involving atomic displacements $OZ$ direction, and the n-plane modes, induced from $E$, in the $OXY$ plane of the sample. In other words, the mechanical band representation can be split into two BRs, $M=M_z\oplus M_{xy}$, with
\begin{eqnarray}\label{MzIrreps}
M_z(\Gamma)&=&\Gamma_{1}^+ (1)+\Gamma_{2}^- (1) \nonumber\\
M_z(K)&=& K_3(2)\\
M_z(M)&=&M_1^+ (1)+ M_2^- (1)\nonumber
\end{eqnarray}
and
\begin{eqnarray}\label{MxyIrreps}
M(\Gamma)&=&\Gamma_{3}^+ (2)+\Gamma_{3}^- (2)\nonumber\\
M(K)&=& K_3(2) + K_1(1) + K_2 (1) \\
M(M)&=&M_1^+ (1)+ M_1^-(1) + M_2^+ (1)+ M_2^- (1).\nonumber
\end{eqnarray}
This facilitates the computation of the symmetry-adapted modes  and clarifies their geometrical nature.

All the irreps at the  $\Gamma$ point have multiplicity one and according to Wigner's theorem the dynamical matrix is fully diagonalized in a basis of symmetry-adapted modes, which therefore are automatically normal modes. The odd-parity (acoustic)  modes are given by
\begin{equation}
\begin{aligned}
& \vec{\varepsilon}_{\mathrm{off}}(\Gamma_2^-)=(0,0,\frac{1}{\sqrt{2}},0,0,\frac{1}{\sqrt{2}}) \\
& \vec{\varepsilon}_{\mathrm{in}}(\Gamma_3^-,1)=(\frac{1}{2},-\frac{i}{2},0,\frac{1}{2},-\frac{i}{2},0) \\ 
& \vec{\varepsilon}_{\mathrm{in}}(\Gamma_3^-,2)=(\frac{1}{2},\frac{i}{2},0,\frac{1}{2},\frac{i}{2},0).
\end{aligned}
\end{equation}
while the optical modes are
\begin{equation}
\begin{aligned}
& \vec{\varepsilon}_{\mathrm{off}}(\Gamma_1^+)=(0,0,\frac{1}{\sqrt{2}},0,0,-\frac{1}{\sqrt{2}}) \\
& \vec{\varepsilon}_{\mathrm{in}}(\Gamma_3^+,1)=(\frac{1}{2},-\frac{i}{2},0,-\frac{1}{2},\frac{i}{2},0) \\
& \vec{\varepsilon}_{\mathrm{in}}(\Gamma_3^+,2)=(\frac{1}{2},\frac{i}{2},0,-\frac{1}{2},-\frac{i}{2},0). \\
\end{aligned}
\end{equation}

The situation changes at the $K$ point, where the irrep $K_3$ has multiplicity two and the corresponding normal modes are linear combinations of the symmetry-adapted modes, while for $K_1$ and $K_2$ the modes are automatically normal. The off-plane modes are given by
\begin{equation}
\begin{aligned}
&\vec{\varepsilon}_{\mathrm{off}}(K_3,1)=(0,0,1,0,0,0) \\
& \vec{\varepsilon}_{\mathrm{off}}(K_3, 2)=(0,0,0,0,0,1). \\
\end{aligned}
\end{equation}
and the in-plane modes by
\begin{equation}
\begin{aligned}
&\vec{\varepsilon}_{\mathrm{in}}(K_1)=(\frac{1}{2},\frac{i}{2},0,\frac{1}{2},-\frac{i}{2},0)\\
&\vec{\varepsilon}_{\mathrm{in}}(K_2)=(\frac{1}{2},\frac{i}{2},0,-\frac{1}{2},\frac{i}{2},0) \\
&\vec{\varepsilon}_{\mathrm{in}}(K_3,1)=(0,0,0,\frac{1}{2},\frac{i}{2},0)\\ 
& \vec{\varepsilon}_{\mathrm{in}}(K_3,2)=(\frac{1}{2},-\frac{i}{2},0,0,0,0).
\end{aligned}
\end{equation}

At the $M$-point only the modes for $M_1^-$ and  $M_2^+$ are automatically normal. The off-plane modes are given by
\\
\begin{equation}
\begin{aligned}
& \vec{\varepsilon}_{\mathrm{off}}(M_1^+)=(0,0,\frac{1}{\sqrt{2}},0,0,\frac{1}{\sqrt{2}}) \\
& \vec{\varepsilon}_{\mathrm{off}}(M_2^-)=(0,0,\frac{1}{\sqrt{2}},0,0,-\frac{1}{\sqrt{2}}), \\
\end{aligned}
\end{equation}
while the off-plane modes are
\begin{equation}
\begin{aligned}
& \vec{\varepsilon}_{\mathrm{in}}(M_1^+)=(\frac{\sqrt{3}}{2\sqrt{2}},\frac{1}{2\sqrt{2}},0,\frac{\sqrt{3}}{2\sqrt{2}},\frac{1}{2\sqrt{2}},0) \\
& \vec{\varepsilon}_{\mathrm{in}}(M_2^-)=(\frac{\sqrt{3}}{2\sqrt{2}},\frac{1}{2\sqrt{2}},0,-\frac{\sqrt{3}}{2\sqrt{2}},-\frac{1}{2\sqrt{2}},0) \\
& \vec{\varepsilon}_{\mathrm{in}}(M_1^-)=(\frac{1}{2\sqrt{2}},-\frac{\sqrt{3}}{2\sqrt{2}},0,-\frac{1}{2\sqrt{2}},\frac{\sqrt{3}}{2\sqrt{2}},0) \\
& \vec{\varepsilon}_{\mathrm{in}}(M_2^+)=(\frac{1}{2\sqrt{2}},-\frac{\sqrt{3}}{2\sqrt{2}},0,\frac{1}{2\sqrt{2}},-\frac{\sqrt{3}}{2\sqrt{2}},0).
\end{aligned}
\end{equation}
\subsection{Eigenvalues of the model dynamical matrix $D(\textbf{k})$}
Changing to the symmetry-adapted basis  turns the dynamical matrix into a block-diagonal form, where the dimension of each block equals the multiplicity of the corresponding irrep. Thus the change of basis yields
the eigenvalues for all the multiplicity one irreps, while  to compute the frequencies associated with a multiplicity two irrep  one has to diagonalize a $2 \times 2$ matrix. At the~$\Gamma$ point the  resulting frequencies depend linearly on the coupling constants

\begin{equation}
\begin{aligned}
& w^2(\Gamma_3^-)= w^2(\Gamma_2^-)=0 \\
& w^2(\Gamma_3^+)= -6(a_1+a_3) \\
& w^2(\Gamma_1^+)= -6(e_1+e_3). \\
\end{aligned}
\end{equation}

This is no longer true at the $K$ and $M$ points, where computing the $K_3$, $M_1^+$ and $M_2^-$ frequencies requires the solution of a quadratic equation due to couplings between off- and in-plane modes. The results are

\begin{widetext}
\begin{equation}
\begin{split}
& w^2(K_1)= -3(a_1+3 a_2+a_3-b_1-b_3+\sqrt{3}d_2) \\
& w^2(K_2)= -3(a_1+3 a_2+a_3+b_1+b_3+\sqrt{3}d_2) ,\\
& w^2(K_3,\pm)=\frac{1}{2} \biggl\{
-3(a_1+3a_2+a_3-\sqrt{3}d_2+e_1+3e_2+e_3)
\pm \biggr[
(-3(a_1+3a_2+a_3-\sqrt{3}d_2-e_1-3e_2-e_3))^2 \\ & \qquad \qquad
+18(h_1+h_3)^{2}
\biggr]^{\frac{1}{2}}
\biggr\} .\\
\end{split}
\end{equation}
and
\begin{equation}
\begin{split}
& w^2(M_1^-)= 2(-2a_1-4a_2+b_1+2b_2), \\
& w^2(M_2^+)= -2(a_1+4a_2+3a_3+b_1-2b_2) ,\\
& w^2(M_1^+,\pm)= -a_1 -4a_2-3a_3+b_1-2b_2-e_1-4e_2-3e_3 
\pm \biggr\{
(-a_1 -4a_2-3a_3+b_1-2b_2+e_1+4e_2+3e_3 )^2 \\ & \qquad \qquad
+4(-h_1+2h_2)^2
\biggr\}^{\frac{1}{2}}, 
\\
& w^2(M_2^-,\pm)= -2a_1-4a_2-b_1-2b_2-2e_1-4e_2 
\pm \biggr\{
(-2a_1-4a_2-b_1-2b_2+2e_1+4e_2)^2
+4(h_1+2h_2)^2
\biggr\}^{\frac{1}{2}}.
\end{split}
\end{equation}
\end{widetext}

\section{\label{app_WL}Wilson loops for the 11 phases}
\begin{figure}[]
\includegraphics[width=8.5cm,keepaspectratio]{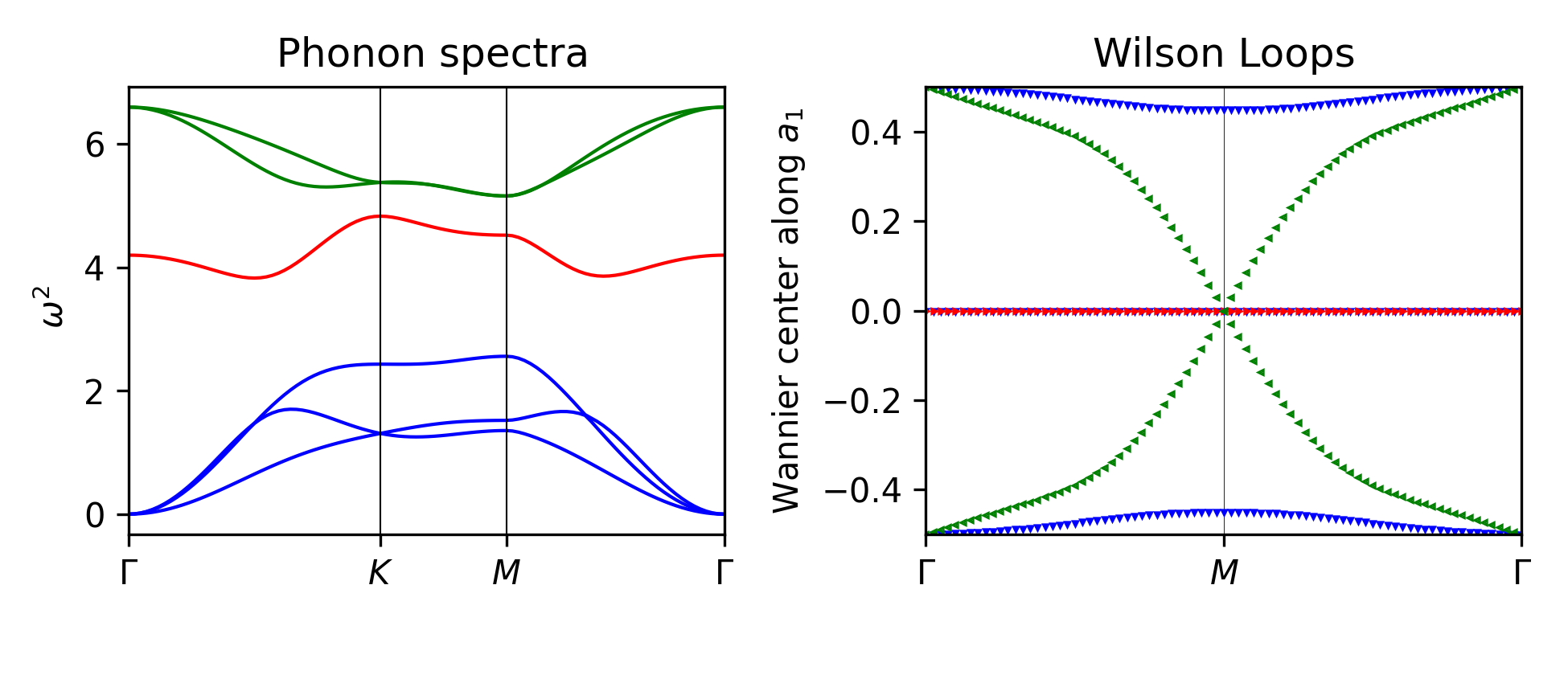}
\caption{\label{fig_phase1} Phase 1 with Winding 1 in the green subset. \\
$a_1 = -1;\ 
e_1 = -0.6;\ 
b_1 =0.3;\ 
h_1 = 0.8;\ 
a_2 = -0.06;\\ 
b_2 = 0.02;\ 
e_2 = -0.06;\ 
g_2 = 0.06;\ 
d_2 = 0.04;\ 
h_2 = 0.04;\\
a_3 = -0.1;\ 
b_3 = 0.1;\ 
e_3 = -0.1;\ 
h_3 = 0.1$.}
\end{figure}

\begin{figure}[]
\includegraphics[width=8.5cm,keepaspectratio]{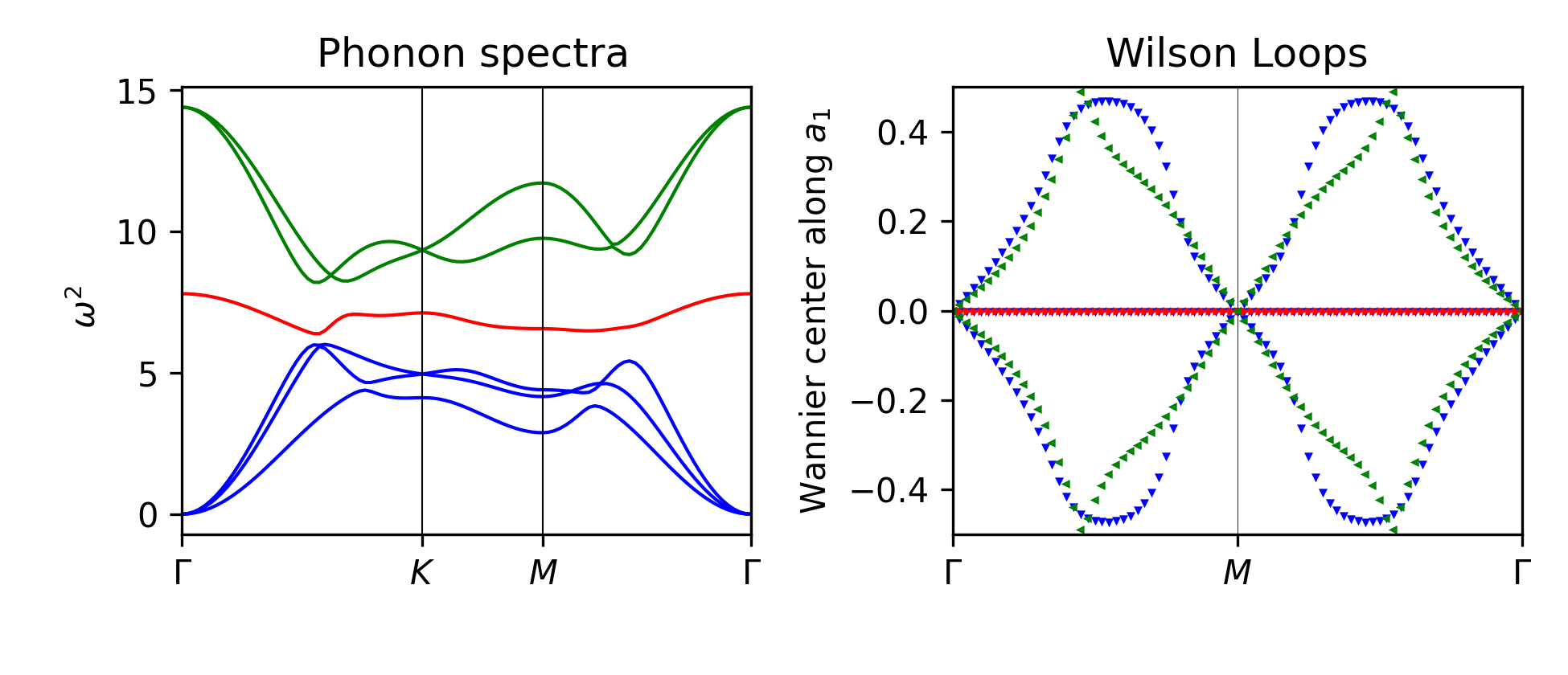}
\caption{\label{fig_phase2} Phase 2 with Winding 2 in the green subset. \\
$a_1 = -1;\ 
e_1 = -0.5;\ 
b_1 = 0.2;\ 
h_1 = 0.2;\ 
a_2 = 0.06.;\\  
b_2 = 0.06;\ 
e_2 = -0.3;\ 
g_2 = 0.2;\ 
d_2 = 0.2;\ 
h_2 = -0.3;\\
a_3 = -1.4;\ 
b_3 = 0.3;\ 
e_3 = -0.8;\ 
h_3 = 0.4$.}
\end{figure}

\begin{figure}[]
\includegraphics[width=8.5cm,keepaspectratio]{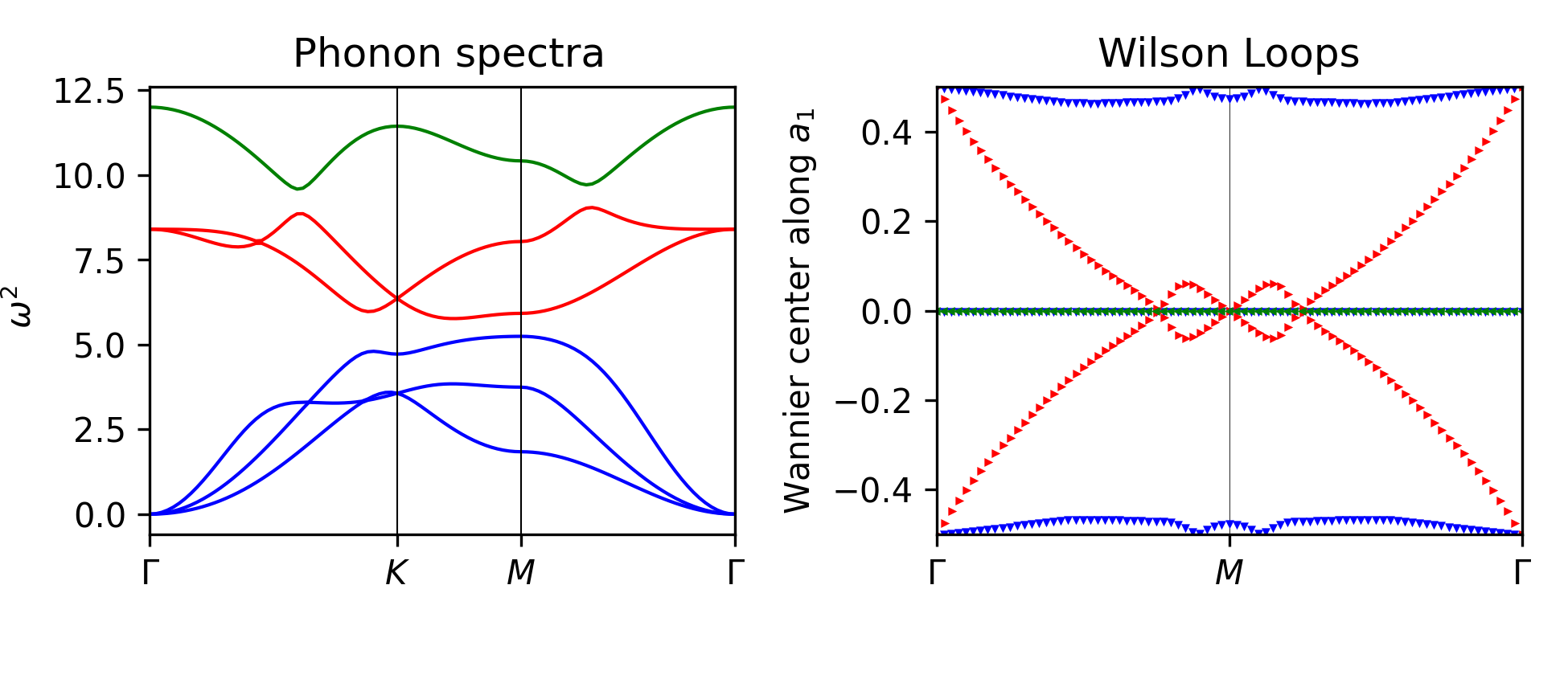}
\caption{\label{fig_phase3} Phase 3 with Winding 1 in the red subset. \\
$a_1 = -1;\
e_1 = -2;\  
b_1 =1.12;\ 
h_1 = 0.4;\ 
a_2 = -0.2;\\ 
b_2 = -0.48;\ 
e_2 = 0;\ 
g_2 = 0;\ 
d_2 = -0.4;\ 
h_2 = -0.12;\\
a_3 = -0.4;\ 
b_3 = 0;\ 
e_3 = 0;\ 
h_3 = 0.04$.}
\end{figure}

\begin{figure}[]
\includegraphics[width=8.5cm,keepaspectratio]{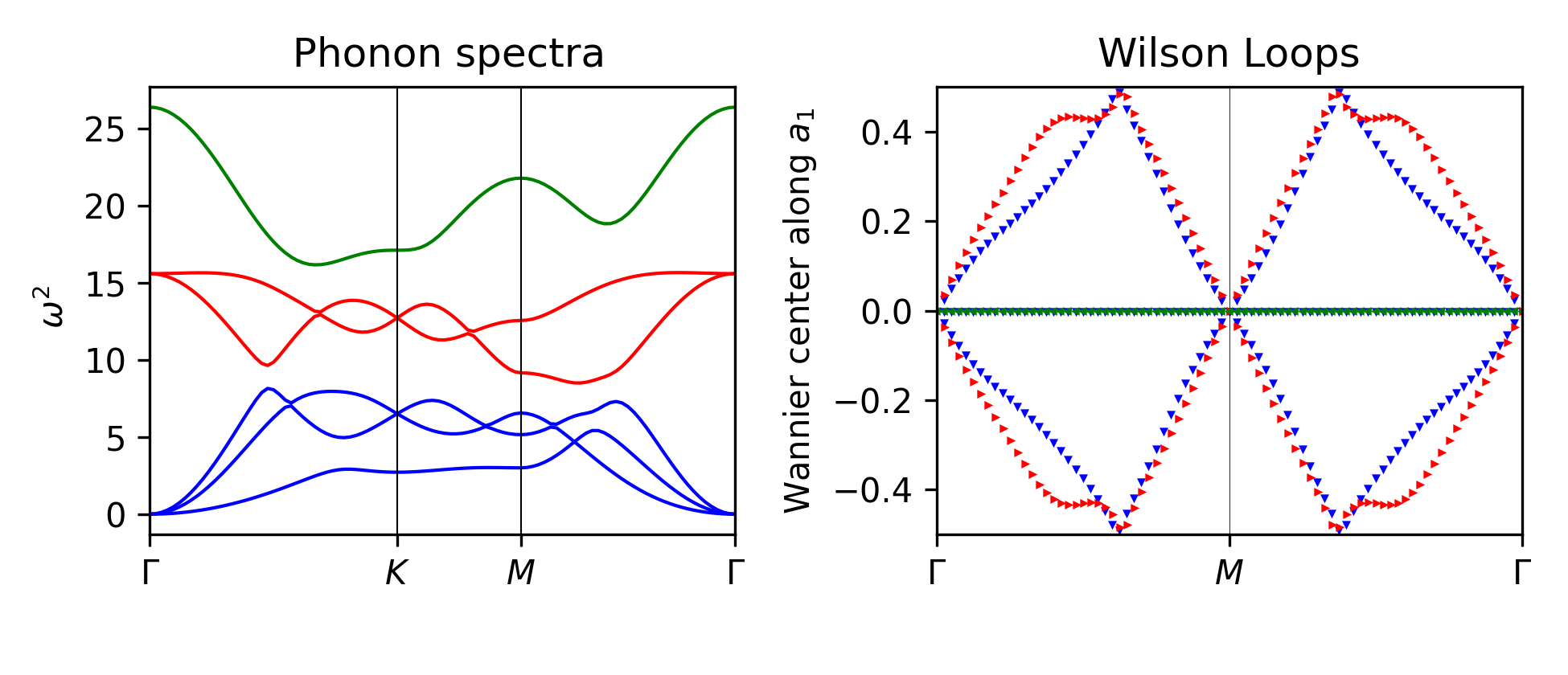}
\caption{\label{fig_phase4} Phase 4 with Winding 2 in red and blue subsets. \\
$a_1 = -1;\ 
e_1 = -1.6;\ 
b_1 = 0.4;\ 
h_1 = 1.6;\ 
a_2 = -0.12;\\ 
b_2 = 0.2;\ 
e_2 = 0.2;\ 
g_2 = -0.4;\ 
d_2 = -0.2;\ 
h_2 = -0.6;\\
a_3 = -1.6;\ 
b_3 = 2;\ 
e_3 = -2.8;\ 
h_3 = -0.4$.}
\end{figure}

\begin{figure}[]
\includegraphics[width=8.5cm,keepaspectratio]{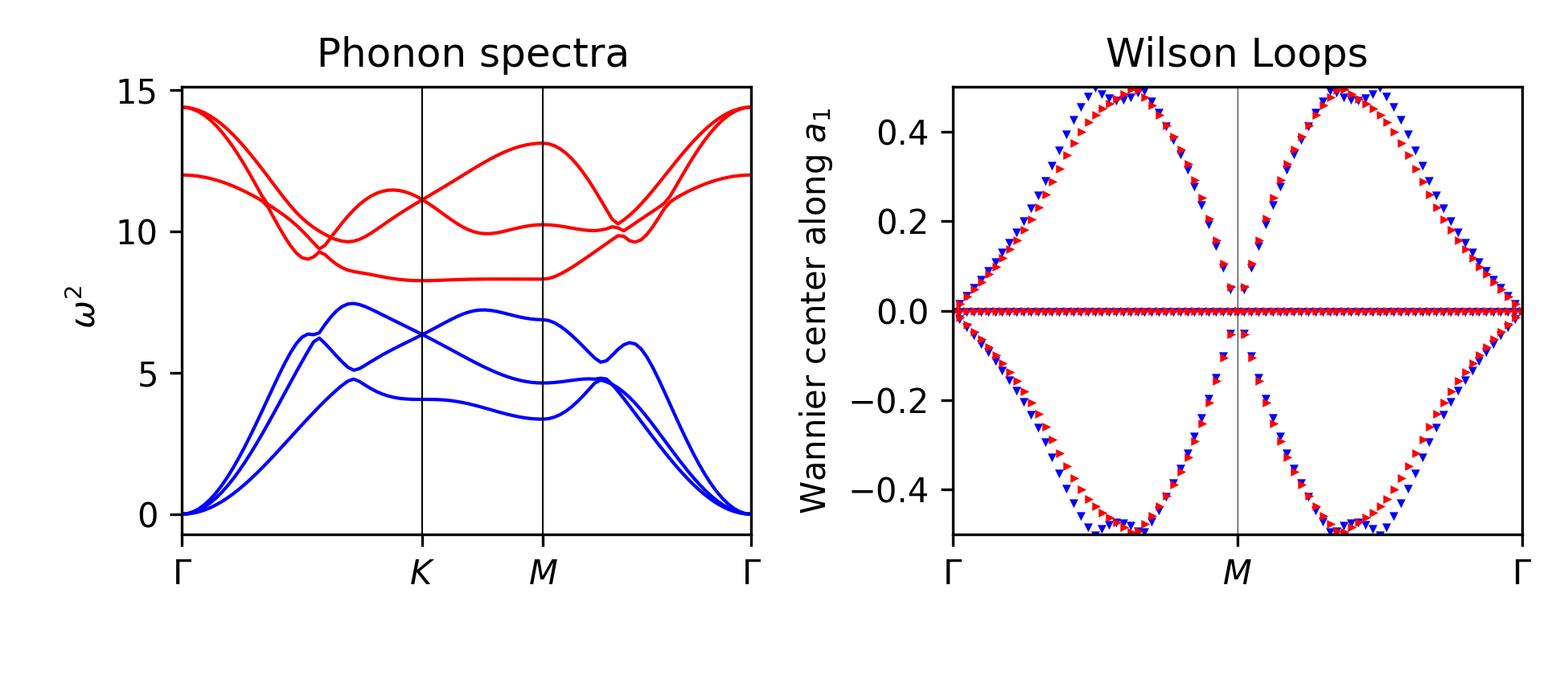}
\caption{\label{fig_phase5} Phase 5 with Winding 2 in the red subset. \\
$a_1 = -1;\ 
e_1 = -1;\ 
b_1 = 0.2;\ 
h_1 = 0.6;\ 
a_2 = 0;\\ 
b_2 = 0.06;\ 
e_2 = -0.36;\ 
g_2 = 0.4;\ 
d_2 = 0.2;\ 
h_2 = -0.3;\\
a_3 = -1.4;\ 
b_3 = 0.5;\ 
e_3 = -1;\ 
h_3 = 0.5$.}
\end{figure}

\begin{figure}[]
\includegraphics[width=8.5cm,keepaspectratio]{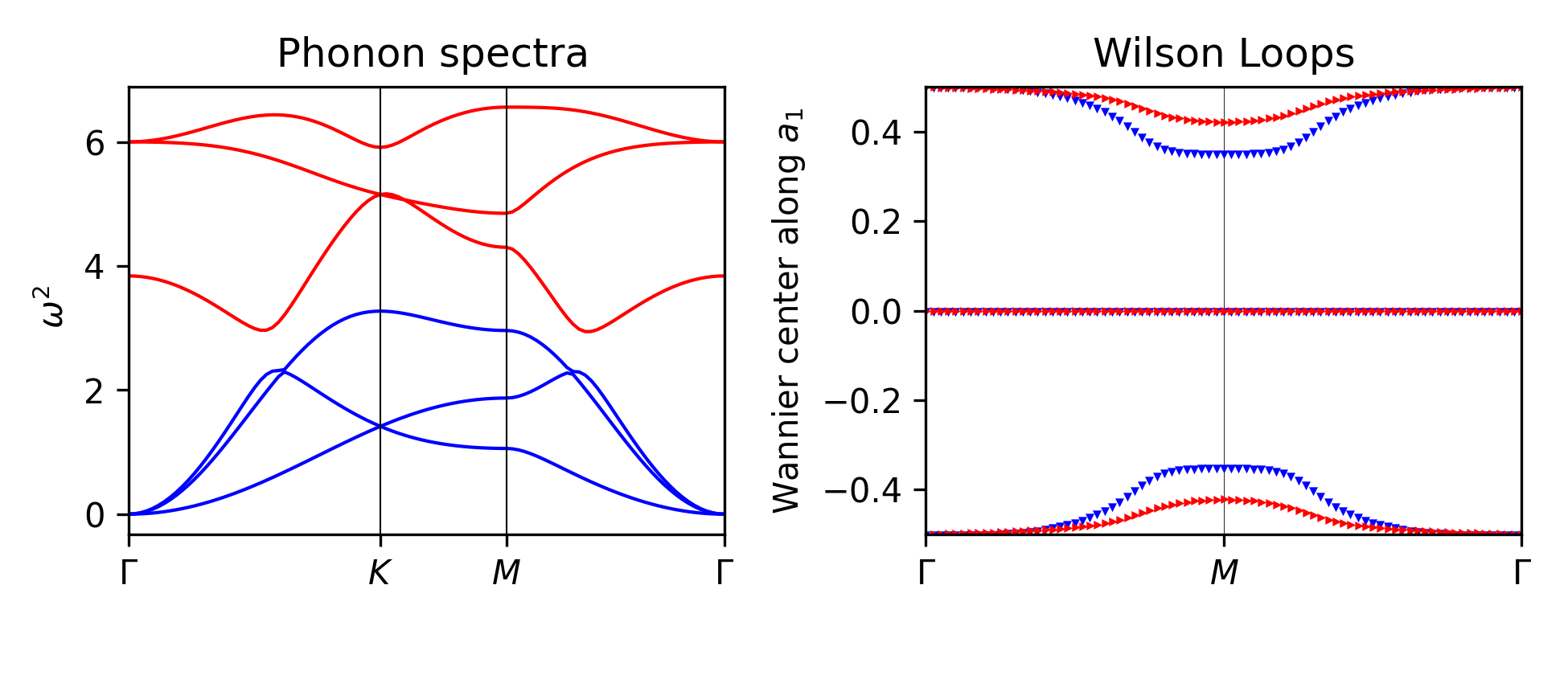}
\caption{\label{fig_phase6} Phase 6 without any winding. \\
$a_1 = -1;\ 
e_1 = -0.6;\ 
b_1 = 0.4;\ 
h_1 = 0.3;\ 
a_2 = -0.2;\\ 
b_2 = 0.04;\ 
e_2 = 0.04;\ 
g_2 = 0.04;\ 
d_2 = 0.04;\ 
h_2 = 0.04;\\
a_3 = 0.;\ 
b_3 = 0.04;\ 
e_3 = -0.04;\ 
h_3 = 0.04$.}
\end{figure}

\begin{figure}[]
\includegraphics[width=8.5cm,keepaspectratio]{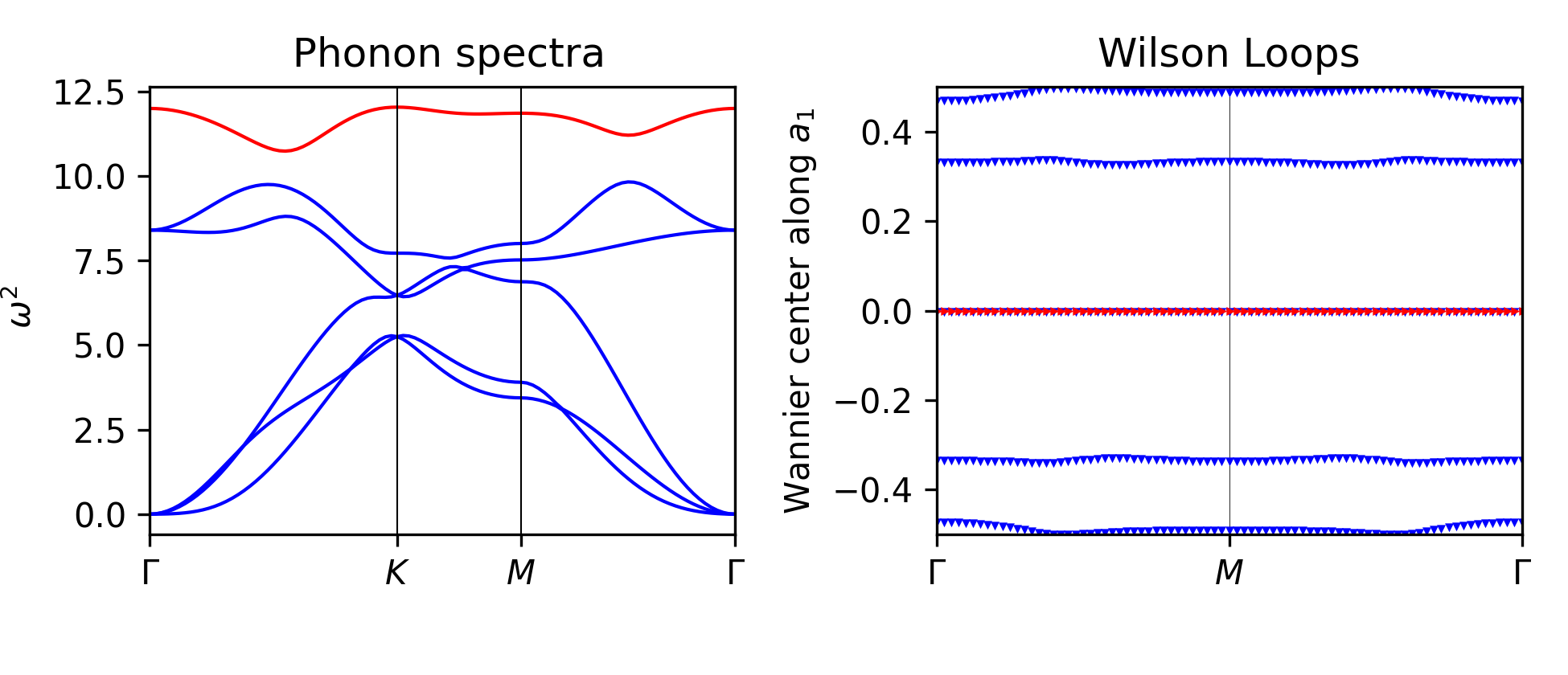}
\caption{\label{fig_phase7} Phase 7 without any winding. \\
$a_1 = -1;\ 
e_1 = -2;\ 
b_1 = 1.12;\ 
h_1 = 0.2;\ 
a_2 = -0.4;\\ 
b_2 = -0.48;\ 
e_2 = 0;\ 
g_2 = -0.4;\
d_2 = -0.4;\ 
h_2 = -0.12;\\ 
a_3 = -0.4;\ 
b_3 = -0.4;\ 
e_3 = 0;\ 
h_3 = 0.08$.}
\end{figure}

\begin{figure}[]
\includegraphics[width=8.5cm,keepaspectratio]{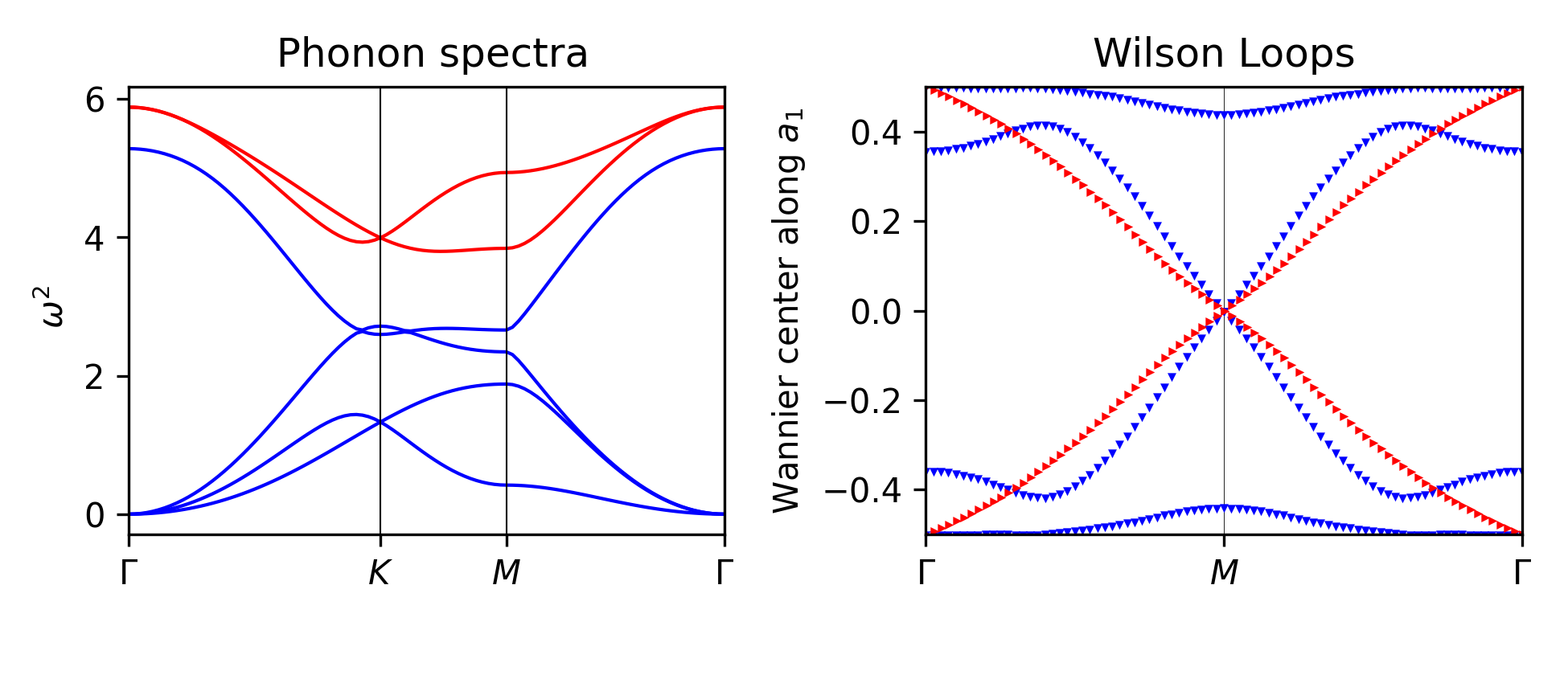}
\caption{\label{fig_phase8} Phase 8 with Winding 1 in the red subset. \\
$a_1 = -1;\ 
e_1 = -0.9;\ 
b_1 = -0.04;\ 
h_1 = 0.6;\ 
a_2 = 0.02;\\ 
b_2 = 0.02;\ 
e_2 = 0.02;\ 
g_2 = 0.02;\ 
d_2 = 0.02;\ 
h_2 = 0.02;\\
a_3 = 0.02;\ 
b_3 = 0.02;\ 
e_3 = 0.02;\ 
h_3 = 0.02$.}
\end{figure}

\begin{figure}[]
\includegraphics[width=8.5cm,keepaspectratio]{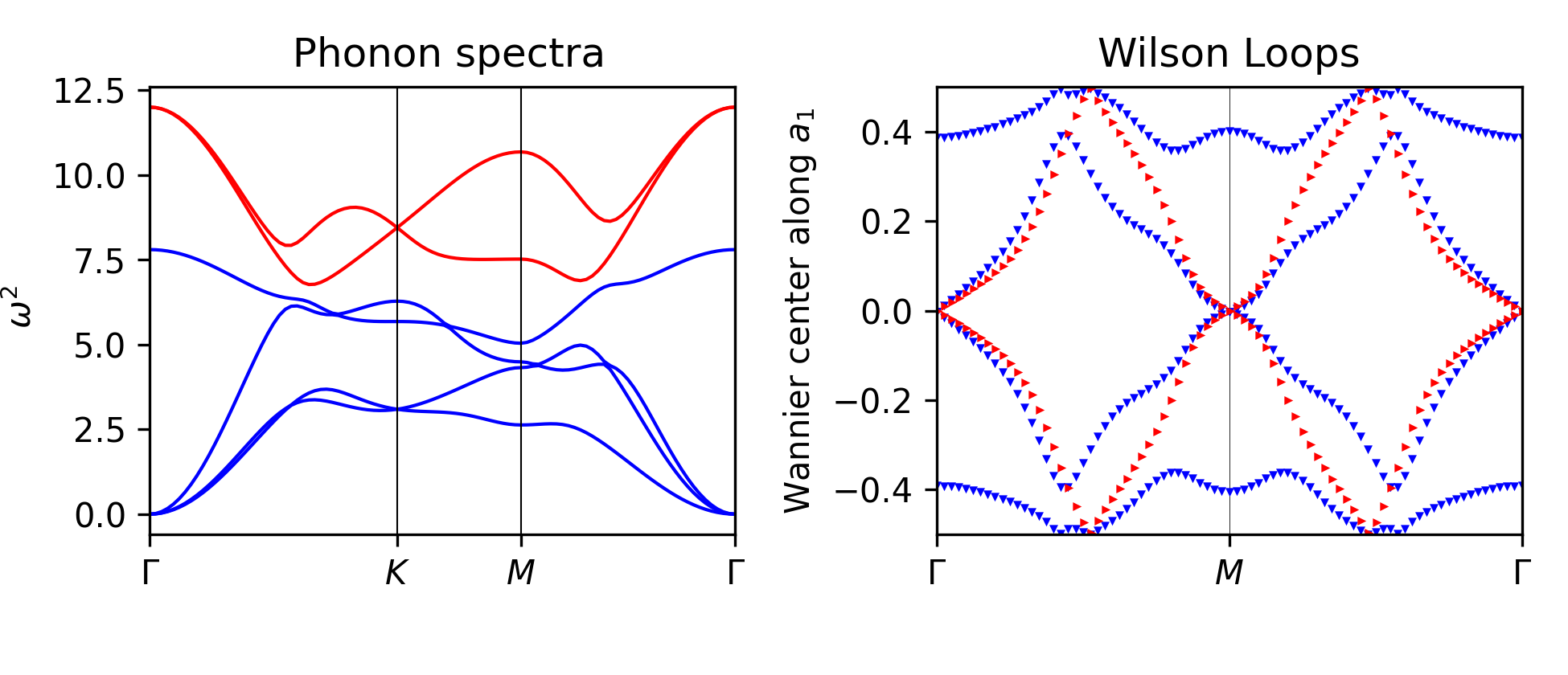}
\caption{\label{fig_phase9} Phase 9 with Winding 2 in the red subset. \\
$a_1 = -1;\ 
e_1 = -0.5;\ 
b_1 = 0.2;\ 
h_1 = 0.8;\ 
a_2 = 0.06;\\ 
b_2 = 0.1;\ 
e_2 = -0.3;\ 
g_2 = 0;\ 
d_2 = -0.1;\ 
h_2 = -0.3;\\
a_3 = -1;\ 
b_3 = -0.3;\ 
e_3 = -0.8;\ 
h_3 = 0.4$.}
\end{figure}

\begin{figure}[]
\includegraphics[width=8.5cm,keepaspectratio]{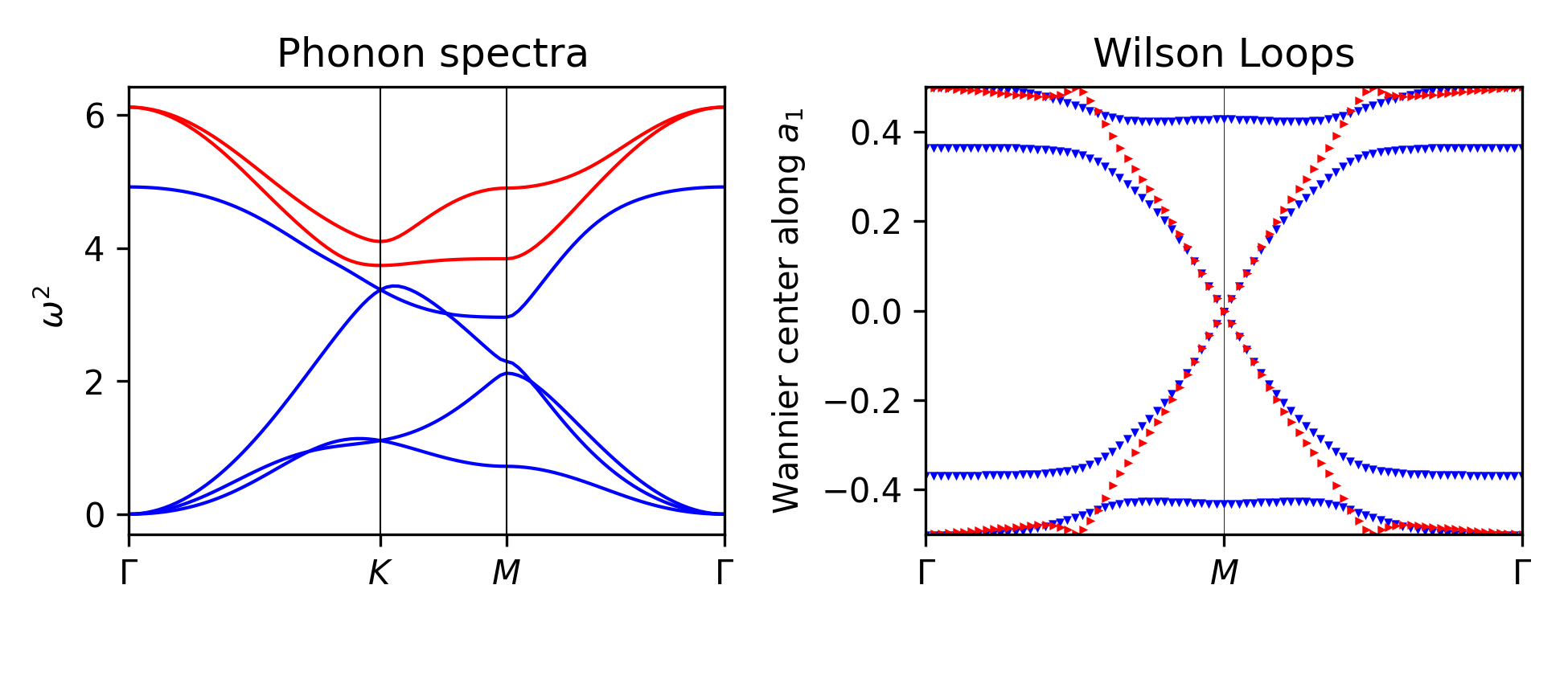}
\caption{\label{fig_phase10} Phase 10 with Winding 1 in the red subset. \\
$a_1 = -1;\ 
e_1 = -0.8;\ 
b_1 = -0.04;\ 
h_1 = 0.6;\ 
a_2 = 0.02;\\
b_2 = 0.02;\ 
e_2 = -0.02;\ 
g_2 = -0.02;\ 
d_2 = -0.2;\ 
h_2 = 0.02;\\
a_3 = -0.02;\ 
b_3 = -0.02;\ 
e_3 = -0.02;\ 
h_3 = -0.1$.}
\end{figure}

\begin{figure}[]
\includegraphics[width=8.5cm,keepaspectratio]{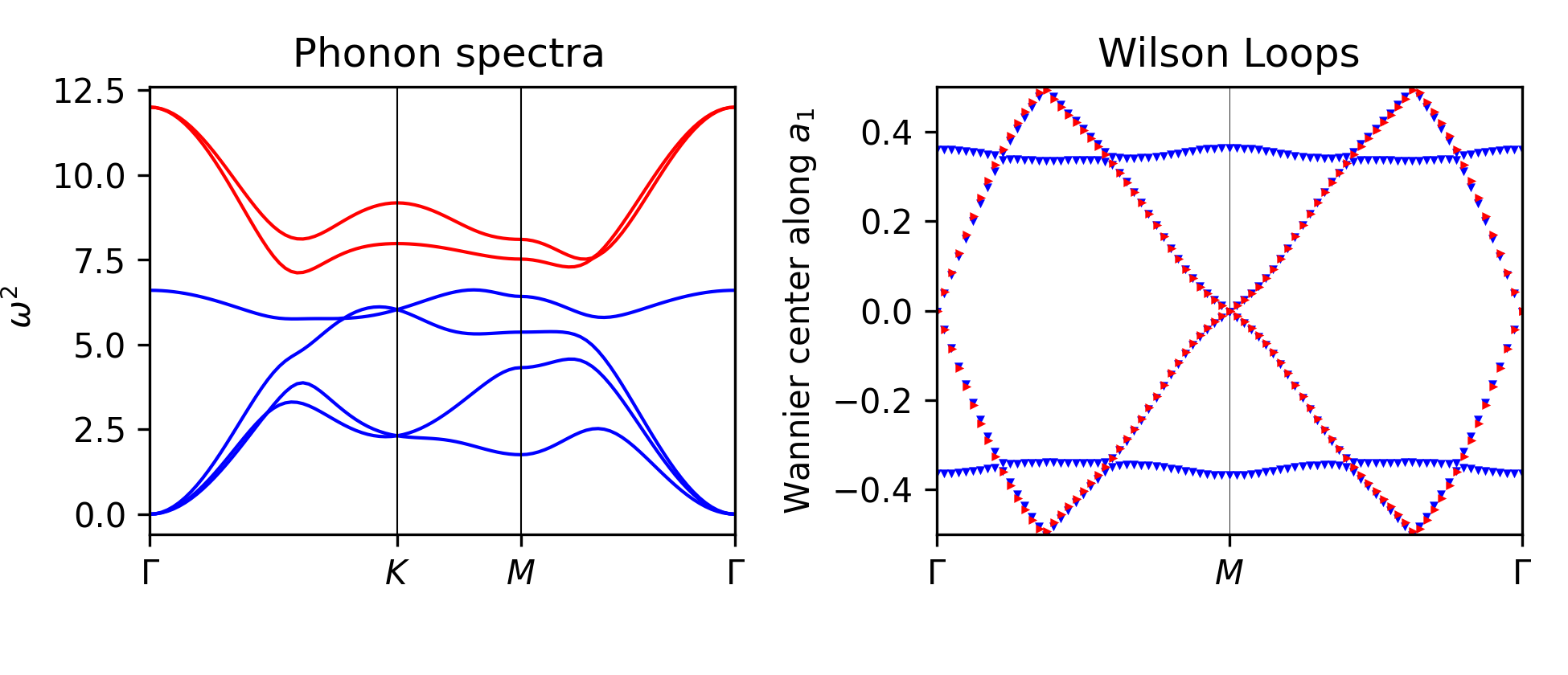}
\caption{\label{fig_phase11} Phase 11 with Winding 2 in red and blue subsets. \\
$a_1 = -1;\ 
e_1 = -0.5;\ 
b_1 = 0.2;\ 
h_1 = -0.2;\ 
a_2 = 0.06;\\
b_2 = 0.1;\ 
e_2 = -0.3;\ 
g_2 = 0;\ 
d_2 = -0.6;\ 
h_2 = -0.3;\\
a_3 = -1;\ 
b_3 = 0;\ 
e_3 = -0.6;\ 
h_3 = 0.04$.}
\end{figure}

The eleven possible phases for the BHL  in Table~\ref{table_EBRs}   were realized within the analytical model by  giving appropriate values to the 14 independent coupling constants $ ({a_1,b_1,h_1,e_1,a_2,b_2,e_2,g_2,d_2,h_2,a_3,b_3,e_3,h_3}) $ defined in Appendix \ref{app_apply_sym} . This enables obtaining a numerical result for the phonon spectra at any point without the need for interpolation. The resulting  phonon bands  were checked for stability (absence of imaginary frequencies) and the existence of gaps separating  the isolated subsets over the whole Brillouin zone, not just along the represented path $\Gamma-K-M-\Gamma$. The phonon bands and Wilson loops are given  in Figs.~\ref{fig_phase1}-\ref{fig_phase11}, while the  windings of the different subsets have been summarized in TABLE \ref{table_wilson}.

\section{\label{app_montecarlo}Mapping DFPT to the analytical model and phase diagram}

\subsection{\label{app_mapping}Mapping DFPT data to the model}

\begin{figure}[h!]
\includegraphics[width=8.5cm,keepaspectratio]{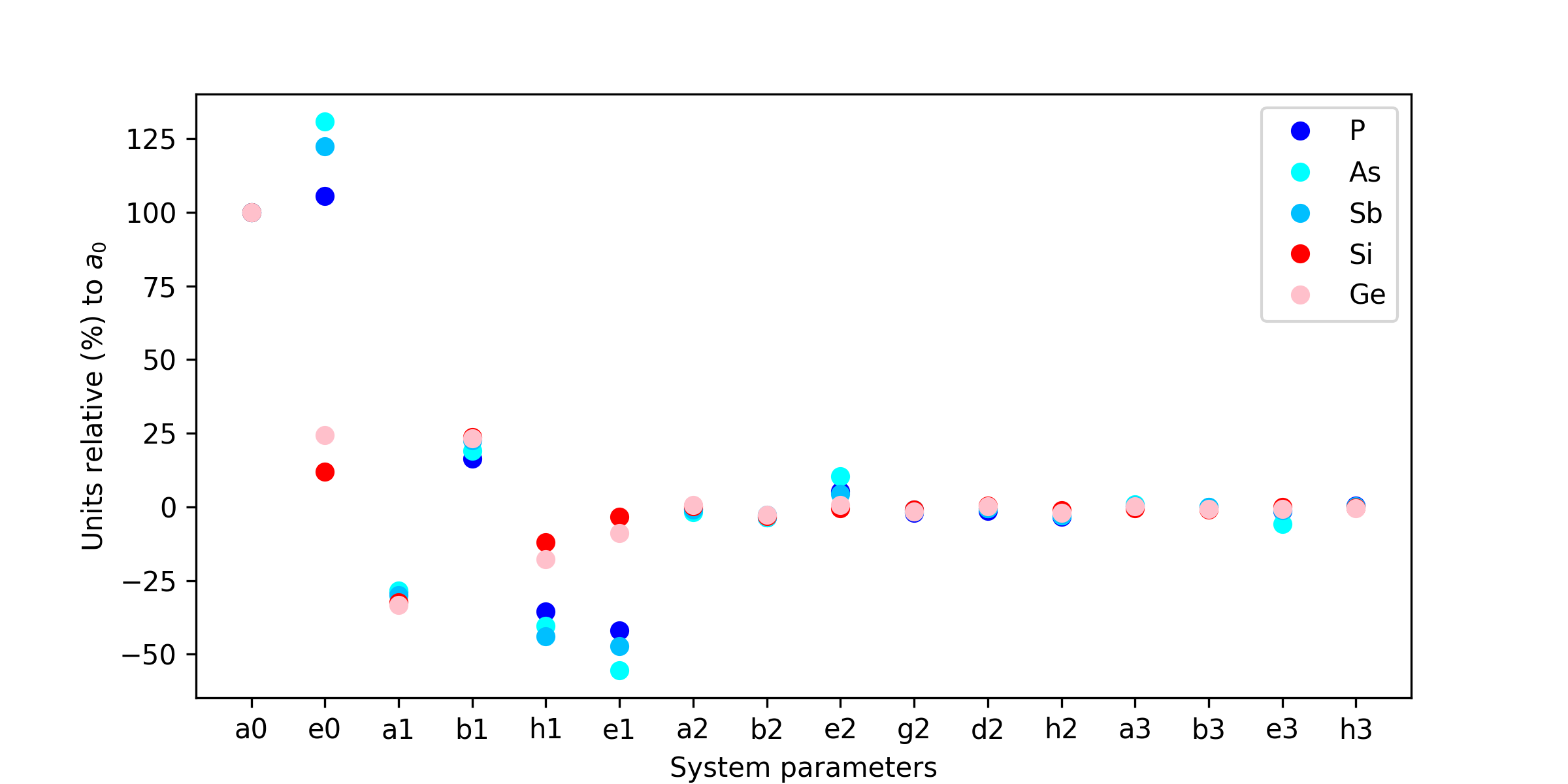}
\caption{\label{fig_system_parameters} Non-zero force constants up to third nearest neighbors as defined in Fig \ref{fig_neighbors}. Dividing the force constants by $a_0$ is equivalent to making a change of units to the squared frequencies and is irrelevant for the topological analysis.}
\end{figure}

Even at first glance  the DFPT phonon bands in Figs.~\ref{fig_phonons} show strong similarities within the families of pentavalent (P,Sb,As) and tetravalent (Si,Ge) materials. Moreover, one can double check this assumption by comparing the force constants extracted by our DFPT calculations. As shown in Fig \ref{fig_system_parameters}, the properly rescaled DFPT force constants defined in Fig \ref{fig_neighbors}  tend to cluster around two points in parameter space. In what follows we will refer to the two  groups as  the P and Si families.

In order to map the DFPT data to the analytical model one should notice that we can describe a material with the BHL symmetries by 14 independent force constants $(a_1,b_1,....,h_3)$, as shown in Appendix \ref{app_apply_sym}. Thus, we can think of these parameters as coordinates in a 14-dimensional space where each point will correspond to a crystal with the BHL structure within our model. 
By inspecting the phonon spectra and irreps in Fig \ref{fig_phonons} one can check that phases 1 and 3 are the closest topological phases to phase 6 for the Si and P families respectively.

\subsection{\label{app_top_neigh} Decay ratios of force constants}

The decay ratios of force constants used in section~\ref{sec_DFPT} and Table~\ref{table_phase_space} are defined by taking the 
mean absolute values of the nonzero  elements of the matrices of force constants in Fig.~\ref{fig_neighbors}. Concretely, if we define
\begin{equation}\label{ratios}
\begin{aligned}
& m_{0}\!=\!\frac{1}{3}\left( 2\left| a_{0}\right| +\left| e_{0}\right| \right)   \\ 
& m_{1}\!=\!\frac{1}{5} \left(\left| a_{1}+b_{1}\right| +\left| a_{1}-b_{1}\right| +\left| e_{1} \right| +2\left| h_{1}\right|\right) \\
&m_{2}\!=\!\frac{1}{9} \left( \left| a_{2}+b_{2} \right| +\left| a_{2}-b_{2} \right|+ \left| e_{2}\right| +2\left| d_{2}\right| +2\left|g_2\right|+2\left| h_{2}\right| \right) \\
& m_{3}\!=\!\frac{1}{5} \left(\left| a_{3}+b_{3}\right| +\left| a_{3}-b_{3}\right| +\left| e_{3} \right| +2\left| h_{3}\right|\right) \ ,
\end{aligned}
\end{equation}
the decay ratios in Table~\ref{table_phase_space} are normalized by $m_0$ and given as  \hbox{$(100:100\, m_1/m_0:  100\, m_2/m_0)$}.

\bibliography{maintextFV}%

\end{document}